\documentclass[showpacs,epj,twocolumn,fontsize=12pt,floatfix]{revtex4-1}
\usepackage{amsmath}
\usepackage[normalem]{ulem}
\usepackage{amsfonts}
\usepackage{amssymb}
\usepackage{amsthm}
\usepackage{graphicx}
\usepackage{color}
\usepackage[utf8]{inputenc}
\usepackage[verbose,hypertexnames=false,bookmarksopenlevel=1,filecolor=blue,
linkcolor=blue,citecolor=blue,pdfstartview=FitH,bookmarksopen,bookmarksnumbered,
colorlinks,plainpages=false,linktocpage]{hyperref}

\definecolor{orange}{rgb}{1,0.5,0}

\DeclareMathOperator\artanh{artanh}
\begin{document}
\title{Entanglement spectra of superconductivity ground states
on the honeycomb lattice}
\author{Sonja Predin}
\email[]
{sonja.predin@physik.uni-regensburg.de}
\author{John Schliemann}
\affiliation{ Institute for Theoretical Physics, University of
Regensburg, D-93040 Regensburg, Germany}

\begin{abstract}
We analytically evaluate the entanglement spectra of the superconductivity 
states in graphene, primarily focusing on the s-wave and 
chiral $ d_{x^{2}-y^{2}}+id_{xy} $ superconductivity states.
We demonstrate that the topology of the entanglement
Hamiltonian can differ from that of the subsystem Hamiltonian.
In particular, the topological properties of the entanglement Hamiltonian
of the chiral $ d_{x^{2}-y^{2}}+id_{xy} $ superconductivity state 
obtained by tracing out one spin direction clearly differ from
those of the time-reversal invariant Hamiltonian of 
noninteracting fermions on the honeycomb lattice.
\end{abstract}
\maketitle
\section{Introduction}
\label{intro}
In graphene, the sixfold symmetry of the honeycomb lattice favours 
the degenerate $ d_{x^{2}-y^{2}} $- and $ d_{xy} $-wave superconductivity 
states. Recent theoretical studies have shown that a s-wave superconductivity 
state \cite{Uchoa2007} and a chiral $ d_{x^{2}-y^{2}} \pm i d_{xy} $ 
superconducting state emerge from electron-electron interactions in
graphene doped to the vicinity of the van-Hove singularity point
\cite{Black-Schaffer07, Honerkamp08, Pathak10, Nandkishore11, Kiesel12, Wang12},
and in lower doped bilayer graphene \cite{Milovanovic12, Vucicevic12, Vafek2014}
(for a recent review, see Ref. \onlinecite{Black-Schaffer14}).
Below the superconducting transition temperature $ T_{C} $, this
degeneracy yields the time-reversal symmetry-breaking
$ d_{x^{2}-y^{2}} \pm i d_{xy} $ state \cite{Platt13, Black-Schaffer14}.

In the past two years, considerable experimental progress has been made regarding the observation
of superconductivity in graphene. 
Evidence of superconductivity has been experimentally observed on 
Ca-intercalated bilayer graphene and graphene laminates
at 4 \cite{Ichinokura2016} and 6.4 K \cite{Chapman2016}, respectively.
Furthermore, additional experimental progress has been made regarding evidence
of superconductivity in Li-decorated monolayer graphene with a transition 
temperature of approximately 5.9 K \cite{Ludbrook2015}.

The discovery of topological phases, which possess topological order 
and cannot be classified by a broken symmetry, has revealed the urgent need for a
tool for characterization of these phases. It has been proven that the 
entanglement entropy obtained from the reduced density matrix can be an 
indicator of the topology in a system\cite{Levin06, Kitaev06, Jiang12}.
Further, Haldane and Li \cite{Li08} have suggested that 
the entanglement spectrum of a system (the full set of eigenvalues of the 
reduced density matrix) contains more information about that system than the 
entanglement entropy, a single number. They have reported a remarkable relationship 
between the excitation spectrum and the edges separating the subsystems, 
considering the entanglement spectrum of the fractional quantum Hall system 
obtained using a spatial cut. It has been suggested that the entanglement 
spectrum constitutes a \textit{tower of states}, which can be regarded as a 
fingerprint of the topological order  \cite{entanglement} 
(for recent reviews, see Refs. \onlinecite{Regnault15, Laflorencie16}). 
The relationship between the entanglement, which can be calculated from 
the ground state, and the edge states, which are excited states of the Hamiltonian
in a sample with boundaries, has been explored in this context.
However, this relationship is not valid in general, as shown in Refs.
\onlinecite{Chandran14, Braganca14, Lundgren14}, in which the various entanglement 
spectra fail to describe the topological phase transitions.

The relationship between the entanglement spectrum obtained by tracing 
out one subsystem and the energy spectrum of the remaining subsystem is attracting 
considerable research attention. Particular focus has been placed on various spin 
ladder systems \cite{Poilblanc10, Cirac11, Peschel11, Lauchli11, Schliemann12, 
Tanaka12, Lundgren13, Chen13, Lundgren16} and on bilayer systems \cite{Schliemann11, 
Schliemann13, Schliemann14}, where a proportionality between the entanglement 
and subsystem Hamiltonians is realized by the strong coupling limit.
However, this relationship is not valid in general, as indicated in
Ref.~\onlinecite{Lundgren12}, in which spin ladders of clearly
nonidentical legs are studied, and in the case of graphene bilayers 
in the presence of trigonal warping \cite{Predin16}.

In a two-dimensional topological superconductor with broken 
time-reversal 
symmetry, the topology can be characterized by a Chern number, 
which is an integral of the 
Berry curvature over the Brillouin zone. The entanglement Chern 
number $C$, i.e., the 
Chern number of the entanglement Hamiltonian obtained from the 
eigenvectors of 
that Hamiltonian, has been suggested to be a topological 
invariant of the 
entanglement Hamiltonian \cite{Predin16, Fukui14, Araki16}.
Note that some investigation of the relationship between the 
energetic and entanglement Hamiltonian topologies has already 
been performed \cite{Predin16}.

In this paper, we present an analytical study of the entanglement spectrum 
of the fermionic ground state on a graphene honeycomb lattice, in the presence of 
superconductivity instability and as obtained by tracing out a single spin direction.
We investigate the relationship between the entanglement and 
energy spectra of the remaining noninteracting part, placing a special focus on the 
correlation between their topologies. 
We show that the entanglement Hamiltonian obtained by tracing out 
one of the subsystems and the Hamiltonian of the remaining subsystem can have completely 
different topologies. This difference is due to the fact that the entanglement 
Hamiltonian is a ground-state property. That is, the $ d_{x^{2}-y^{2}}+id_{xy} $ 
superconductivity state breaks the time-reversal symmetry of the 
superconductivity Hamiltonian; this behavior is reflected in the ground state of the 
composite superconductivity Hamiltonian. Further,
the entanglement Hamiltonian is constructed from that ground state.

This paper is organized as follows: In Section \ref{model}, we introduce
the model Hamiltonian and discuss the different superconductivity paired
states that can arise on the honeycomb lattice.  
Classification of the topological phases
of the superconductivity states on the honeycomb lattice based on their different 
symmetries is also performed in this section. The entanglement spectrum 
obtained from the Bardeen-Cooper-Schrieffer 
ground state by tracing out a single spin direction is analyzed in Section
\ref{entanglement}.  
Our primary interest in this section is to explore the relationship between the geometrical 
and topological properties of the entanglement 
Hamiltonian and the remaining noninteracting Hamiltonian. 
We also discuss the case of
sublattices B are traced out.
We close with a summary and an overview of the future research outlook, which is presented in Section \ref{conclusion}.
Some technical details on the analytical derivation of the full eigenstates of 
the noninteracting fermionic system on the honeycomb lattice
in the presence of superconductivity instabilities are presented, along with correlation matrix calculations, in Appendices \ref{app1} and \ref{app2}. 
\section{Model Hamiltonian}
\label{model}

\begin{figure}[t]
\includegraphics[width=0.7\columnwidth]{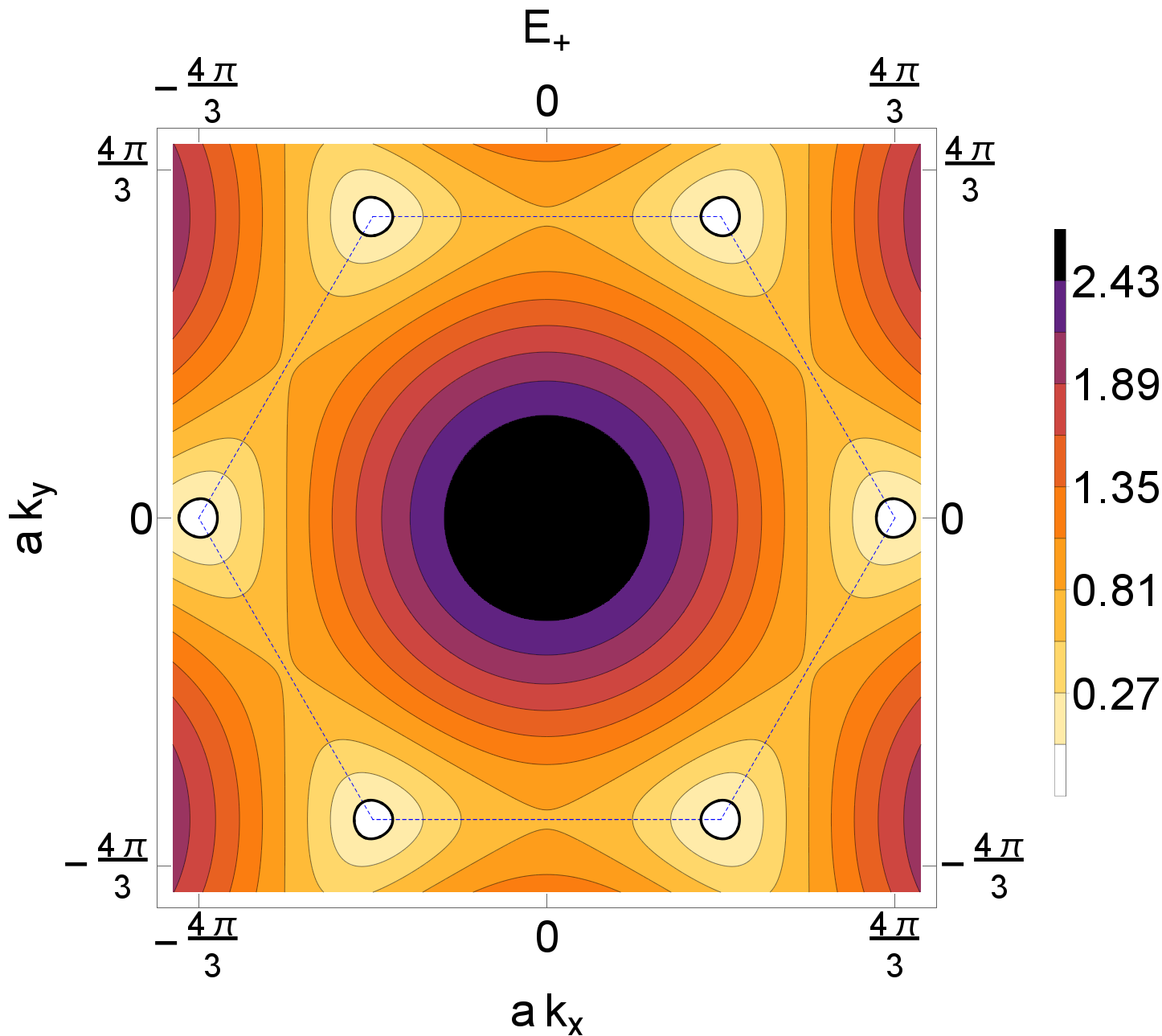}
\includegraphics[width=0.7\columnwidth]{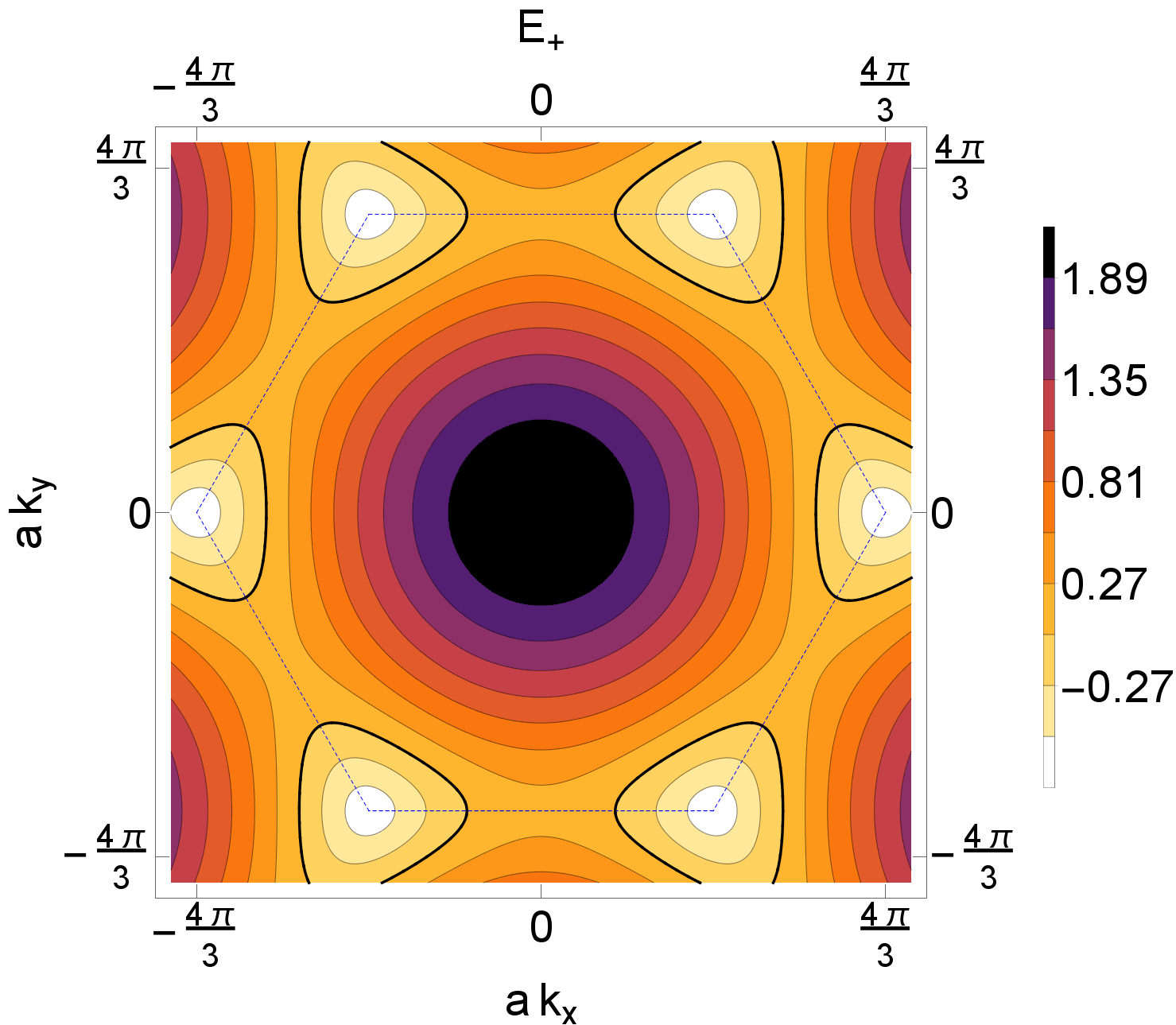}
\includegraphics[width=0.7\columnwidth]{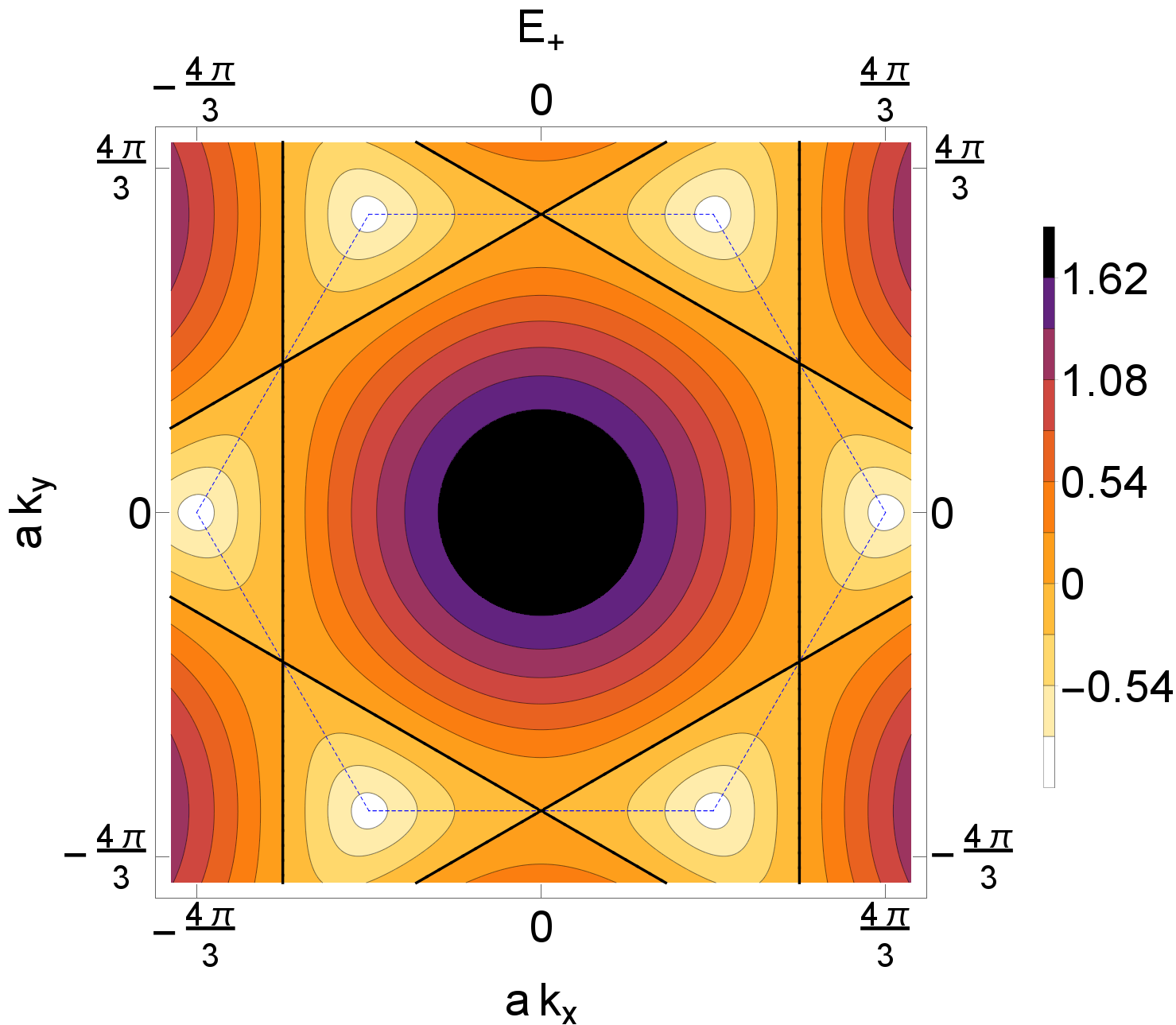}
\caption{(Color Online)
Brillouin zone with density plot of $|\gamma(\vec k)|-\frac{\mu}{t}$ for:
(a) $ \frac{\mu}{t}=0.2 $; (b)  $ \frac{\mu}{t}=0.8 $; and 
(c) $ \frac{\mu}{t}=1 $. The edge of the first Brillouin zone is marked by dashed
blue lines.}
\label{fig1}
\end{figure}

The tight-binding Hamiltonian for free fermions on a graphene honeycomb lattice with a 
single $ 2p_{z} $ orbital per carbon (C) atom is
\begin{align}
H_{0}=&-t\sum_{\langle ij \rangle}\sum_{\sigma=\uparrow,\downarrow}
\left(a_{i,\sigma}^{\dagger}b_{j,\sigma}+h.c.\right) \nonumber \\
&-\mu\sum_{i,\sigma}\left(a_{i,\sigma}^{\dagger}a_{i,\sigma}+
b_{i,\sigma}^{\dagger}b_{i,\sigma}\right), \label{non-interacting}
\end{align}
where $ t $ is the hopping energy between the nearest-neighbor C atoms,
$ \mu $ is the chemical potential and
$ a_{i,\sigma} $ ($ a_{i,\sigma}^{\dagger} $) and $ b_{i,\sigma} $ ($ b_{i,\sigma}^{\dagger} $) are the onsite annihilation 
(creation) operators for electrons on sublattices A and B, respectively, with spin 
$ \sigma=\uparrow,\downarrow $.
Diagonalization of Eq.~(\ref{non-interacting}) yields the  energy spectrum $ \pm E_{\pm} $, with
\begin{equation}
E_{\pm}=\pm t \vert \gamma(\vec{k})\vert - \mu,
\end{equation}
where $ \gamma(\vec{k})=\sum_{\vec{\delta}}\exp\left(i\vec{k}\cdot\vec{\delta}\right)$ and
$ \vec{\delta} $ is a nearest-neighbor vector.
In what follows, we use coordinates with
\begin{align}
& \vec\delta_{1}=a\left(0,\frac{1}{\sqrt{3}}\right), \\
& \vec\delta_{2,3}=\frac{a}{2}\left(\pm 1, -\frac{1}{\sqrt{3}}\right), 
\end{align}
where $a=1.42$~\AA\, is the distance between neighboring C atoms,
such that the two inequivalent corners of the first Brillouin zone can
be expressed as
\begin{equation}
{\vec K}_{\pm}=\pm \left(\frac{4\pi}{3a},0\right).
\label{defK}
\end{equation}
The energy spectrum of the free fermions over 
the first Brillouin zone is visualized in Fig.~\ref{fig1}. 

In order to apply the mean-field approximation, we define the
superconductivity order parameter as a three-component complex vector
\begin{eqnarray}
\overrightarrow{\Delta}\equiv \left(\Delta_{\vec{\delta}_{1}},\Delta_{\vec{\delta}
_{2}},
\Delta_{\vec{\delta}_{3}}\right),
\end{eqnarray}
where the components are defined by
\begin{eqnarray}
\Delta_{\vec{\delta}}=\left\langle 
a_{i\uparrow}b_{i+\vec{\delta}\downarrow} - 
a_{i\downarrow}b_{i+\vec{\delta}\uparrow} \right\rangle. 
\end{eqnarray}
We study the superconductivity pairing arising from the nearest-neighbor attractive interaction
\begin{align}
H_{int}=\sum_{i,\vec{\delta}}\Delta_{\vec{\delta}}\left(a_{i\uparrow}^{\dagger}
b_{i+\vec{\delta}\downarrow}^{\dagger} - 
a_{i\downarrow}^{\dagger}b_{i+\vec{\delta}\uparrow}^{\dagger}\right),
\end{align}
with the limit of strong onsite interaction.
The resulting mean-field Hamiltonian can be expressed in momentum space as
\begin{align}
H_{MF}=&-t\sum_{\vec{k}\sigma}\left(\gamma(\vec{k})a_{\vec{k}\sigma}^{\dagger}b_{\vec{k}\sigma}+h.c.\right) \nonumber\\
&-\mu\sum_{\vec{k}\sigma}\left(a_{\vec{k}\sigma}^{\dagger}a_{\vec{k}\sigma} 
+b_{\vec{k}\sigma}^{\dagger}b_{\vec{k}\sigma}\right)\nonumber \\
&-J \sum_{\vec{k},\vec{\delta}}\left(\Delta_{\vec{\delta}}e^{i\vec{k}\vec{\delta}}
\left(a_{\vec{k}\uparrow}^{\dagger}b_{-\vec{k}\downarrow}^{\dagger}-a_{\vec{k}\downarrow}^{\dagger}b_{-\vec{k}\uparrow}^{\dagger}\right)+h.c.\right), \label{HamMF}
\end{align} 
where $J$ is the effective pairing potential arising from the electron-electron interaction.
The kinetic part of the previous 
Hamiltonian can be diagonalized by introducing the following
transformations
\begin{eqnarray}
c_{\vec{k},\sigma}=\frac{1}{\sqrt{2}}(a_{\vec{k},\sigma}-e^{i \cdot \phi_{\vec{k}}}b_{\vec{k},\sigma}) 
,\nonumber \\
 d_{\vec{k},\sigma}=\frac{1}{\sqrt{2}}(a_{\vec{k},\sigma}+e^{i \cdot \phi_{\vec{k}}}b_{\vec{k},\sigma}),
\end{eqnarray} 
where the phase $ \phi_{\vec{k}} $ is defined as $ \phi_{\vec{k}}=\arg 
(\gamma_{\vec{k}}) $. 
Note that $ c_{\vec{k}, \vec{\sigma}}^{\dagger} $ and $ d_{\vec{k},\vec{\sigma}}^{\dagger} $ create an electron in 
the upper and lower Bogoliubov bands, respectively.

Thus, introducing the energy
basis, the Hamiltonian becomes
\begin{align}
& H_{MF} = - t \sum_{\vec{k},\sigma} \vert \gamma_{\vec{k}} \vert(d_{\vec{k},
\sigma}^{\dagger}d_{\vec{k},\sigma}-
c_{\vec{k},\sigma}^{\dagger}c_{\vec{k},\sigma}) \nonumber \\
&- \mu \sum_{\vec{k}, \sigma} (d_{\vec{k},
\sigma}^{\dagger}d_{\vec{k},\sigma}+c_{\vec{k},
\sigma}^{\dagger}c_{\vec{k},\sigma}) \nonumber \\
&-  J \sum_{\vec{k}} \sum_{\vec{\delta}} 
\left(\Delta_{\vec{\delta}}\left(\cos(\vec{k}\vec{\delta} - 
\phi_{\vec{k}})(d_{\vec{k},
\uparrow}^{\dagger}d_{-\vec{k},\downarrow}^{\dagger} - c_{\vec{k},
\uparrow}^{\dagger}c_{-\vec{k},\downarrow}^{\dagger}) \right. \right. \nonumber \\
&+\left. \left. i \sin(\vec{k}\vec{\delta} - 
\phi_{\vec{k}})(c_{\vec{k},
\uparrow}^{\dagger}d_{-\vec{k},\downarrow}^{\dagger} - d_{\vec{k},
\uparrow}^{\dagger}c_{-\vec{k},\downarrow}^{\dagger})\right) + h.c. \right). \label{Hamil}
\end{align} 
The third line in this Hamiltonian is the intraband pairing, 
containing an order parameter that is even in k-space
and corresponding to the spin-singlet pairing. The fourth line  
is the interband pairing, 
containing an order parameter that is odd in k-space
and corresponding to the spin-triplet pairing. 
We use the definitions 
\begin{eqnarray}
C_{\vec{k}} = J \sum_{\vec{\delta}} 
\Delta_{\vec{\delta}}\cos(\vec{k}\vec{\delta} - 
\phi_{\vec{k}}),
\end{eqnarray} and 
\begin{eqnarray}
S_{\vec{k}} = J 
\sum_{\vec{\delta}} 
\Delta_{\vec{\delta}}\sin(\vec{k}\vec{\delta} - 
\phi_{\vec{k}}).
\end{eqnarray}
The corresponding span of the superconducting order 
parameter is 
\begin{equation}
\overrightarrow{\Delta}=\left\lbrace 
\begin{array}{lll}
\Delta(1,1,1), \\
\Delta(2,-1,-1), \\
\Delta(0,-1,1), \\
\end{array}
\right.
\end{equation}
where $ \Delta $ is the self-consistent superconductivity order parameter.
In what follows, we use the redefinition $ J\Delta\equiv \Delta $.
The linearized self-consistence equations of the order parameter are invariant
with respect to the hexagonal group $ C_{6v} $ \cite{Black-Schaffer07}, i.e.,
the symmetry group of the honeycomb lattice.
The first solution corresponds to the s-wave, $  \overrightarrow{\Delta} = \Delta(1,1,1) $, 
belonging to the natural A1 irreducible representation of the $ C_{6v} $ 
group of the honeycomb lattice. The A1 irreducible representation is spanned 
by the vector
$ \vec{u}_{1}=(1,1,1) $. The final two solutions,$\overrightarrow{\Delta} = \Delta(2,-1,-1)$ and 
$\overrightarrow{\Delta} = \Delta(0,-1,1)$, belong to the
two-dimensional subspace of the $ S_{3} $ group \cite{Poletti11}, the span of which is 
$\vec{u}_{2}=(2,-1,-1)$ and $\vec{u}_{3}=(0,-1,1)$.
The second (corresponding to the $ d_{x^{2}-y^{2}} $
wave) and third (corresponding to the $ d_{xy} $ wave) 
solutions belong to the E1 and E2 irreducible representations of the $ S_{3} $ group, respectively.
From the symmetry perspective, it is noteworthy that 
every combination of the $ d_{x^{2}-y^{2}} $
and $ d_{xy} $ waves is possible. However, it has been shown that the 
$ d_{x^{2}-y^{2}} \pm i d_{xy} $-wave superconductivity state with an 
order parameter
\begin{equation}
\overrightarrow{\Delta}_{d_{x^{2}-y^{2}} \pm i d_{xy} }=\frac{1}{\sqrt{3}}\Delta\left(
\begin{array}{lll}
1\\
e^{\pm \frac{2i\pi}{3}} \\
e^{\mp \frac{2i\pi}{3}}
\end{array}\right),
\end{equation}
is preferred in graphene below $ T_{C} $ for a 
superconductivity coupling strength $J$ that is not excessively large, 
and for doping up to
and in the vicinity of the van-Hove singularity point \cite{Black-Schaffer07}.

The s-wave superconductivity order parameter is given by
$ \Delta(\vec{k})=\gamma(\vec{k}) $, while the $ d_{x^{2}-y^{2}} +i d_{xy} $-wave superconductivity order parameter is
\begin{equation}
\Delta_{d\pm id}(\vec{k})=\cos\left(\frac{\pi}{3}\right)\Delta_{d_{x^{2}-y^{2}}}(\vec{k})\pm 
\sin\left(\frac{\pi}{3}\right)\Delta_{d_{xy}}(\vec{k}),
\label{def_order_d}
\end{equation}
with
\begin{align}
&\Delta_{d_{x^{2}-y^{2}}}(\vec{k})=2\Delta\left(e^{iak_{x}}-e^{-i\frac{a}
{2}k_{x}}\cos(\frac{a\sqrt{3}}{2}k_{y})\right), \label{def_order_d1} \\
&\Delta_{d_{xy}}(\vec{k})=-2i\Delta \sin\left(\frac{a\sqrt{3}}{2}k_{y}\right)e^{-i\frac{a}
{2}k_{x}}. \label{def_order_d2}
\end{align}

Introducing the spinor 
\begin{align}
\varphi_{\vec{k}}^{\dagger}=\left(a_{\vec{k}\uparrow}^{\dagger}, 
b_{\vec{k}\uparrow}^{\dagger}, a_{\vec{k}\downarrow}^{\dagger}, 
b_{\vec{k}\downarrow}^{\dagger},a_{-\vec{k}\uparrow}, b_{-\vec{k}\uparrow},
 a_{-\vec{k}\downarrow}, b_{-\vec{k}\downarrow}
\right),
\end{align}
the Hamiltonian of Eq.~(\ref{HamMF}) can be expressed as
\begin{equation}
H_{MF}=\frac{1}{2}\sum_{\vec{k}}\varphi_{\vec{k}}^{\dagger}\mathcal{M}_{\vec{k}}
\varphi_{\vec{k}},
\end{equation}
where
\begin{align}
\mathcal{M}_{\vec{k}}=\left(\begin{array}{cccc}
\zeta(\vec{k}) & 0 & 0 & -\overline{\Delta} (\vec{k}) \\
0 & \zeta(\vec{k}) & \overline{\Delta} (\vec{k}) & 0 \\
0 & \overline{\Delta}^{*}(-\vec{k}) & -\zeta^{*}(-\vec{k}) & 0 \\
- \overline{\Delta}^{*}(-\vec{k}) & 0 & 0 & -\zeta^{*}(-\vec{k})
\end{array}\right), \label{mat_ham}
\end{align}
with
\begin{align}
\zeta(\vec{k})=\left(\begin{array}{cc}
-\mu & -t \gamma(\vec{k}) \\
-t \gamma^{\ast}(\vec{k}) & -\mu
\end{array}\right), \\
\overline{\Delta}(\vec{k})=\left(\begin{array}{cc}
0 & \Delta(\vec{k}) \\
\Delta(-\vec{k}) & 0
\end{array}\right).
\end{align}

The resultant Hamiltonian indicates that the spin-singlet superconductivity state
without spin-orbit coupling is invariant under the spin SU(2) rotation. Hence, we obtain the condition
\begin{align}
\left[J_{i},\mathcal{M}(\vec{k})\right]=0, \qquad 
J_{i}=\left(\begin{array}{cc}
s_{i} & 0 \\
0 & -s_{i}^{\ast}
\end{array}\right), \qquad
(i=x,y,z).
\end{align}
As a result of the spin SU(2) rotation, it is sufficient to use the spinor 
$ \Psi_{\vec{k}}^{\dagger}=
(a_{\vec{k}\uparrow}^{\dagger}, b_{\vec{k}\uparrow}^{\dagger}, 
a_{-\vec{k}\downarrow}, b_{-\vec{k}\downarrow}) $ 
in order to express the Hamiltonian of the superconductivity state on the 
honeycomb lattice in the form
\begin{eqnarray}
H_{MF}=\sum_{\vec{k}}\Psi^{\dagger}_{\vec{k}}h(\vec{k})\Psi_{\vec{k}},
\end{eqnarray}
where
\begin{equation}
h(\vec{k})=\left(\begin{array}{cccc}
-\mu & -t\gamma(\vec{k}) & 0 & -\Delta(\vec{k})\\
-t \gamma^{\ast}(\vec{k}) & - \mu & -\Delta(-\vec{k}) & 0\\
0 & - \Delta^{\ast}(-\vec{k}) & \mu & t\gamma^{\ast}(-\vec{k})\\
-\Delta^{\ast}(\vec{k}) & 0 & t\gamma(-\vec{k}) & \mu
\end{array}\right). \label{HamS}
\end{equation}
When the superconductivity order parameter is pure real the Hamiltonian
$ h(\vec{k}) $ satisfies
\begin{align}
T h(\vec{k}) T^{-1}=h(-\vec{k}), \label{time}
\end{align}
where $ T=K $ mimics time-reversal symmetry. 
The condition given in Eq.~(\ref{time}) can 
satisfy a real superconductivity order parameter only.
The $ d_{x^{2}-y^{2}} +i d_{xy} $-wave superconductivity order parameter given by
Eq.~(\ref{def_order_d}) breaks the time-reversal symmetry.
It appertains to the 
CI-class in the Altland-Zirnbauer classification of the topological insulators and 
superconductors \cite{Sato, Altland97, Schnyder08}.
Furthermore, it is possible to classify two-dimensional C-class superconductors 
using the Chern number $C$. Note that the nontrivial topology of the 
$ d_{x^{2}-y^{2}} +i d_{xy} $-wave 
superconductivity state is denoted by the Chern number $C=2$.
\section{Entanglement Spectra}
\label{entanglement}
A method for analytically calculating the entanglement spectrum of a
free-fermion system is given in Refs. \onlinecite{Peschel03, Cheong04, 
Schliemann13}. Here, we generalize this method to superconductivity
systems, using an approach similar to that described in Refs. \onlinecite{Borchmann14, Kim14}.

The entanglement Hamiltonian can be constructed as a single-particle operator in a quadratic matrix \cite{Peschel03, Cheong04, Schliemann13}, as it is
completely determined by any correlation matrix of operators acting 
on the remaining part after the subsystem has been traced out. Our system consists of two subsystems, A and B. 
The reduced density matrix for 
subsystem A, defined as $ \rho_{A}={\rm tr}_{B}\rho $, can be formulated as in the 
free fermion case, such that $ \rho_{A}=\frac{1}{Z}e^{-H_{{\rm ent}}}  $,
using the entanglement spectrum $ H_{{\rm ent}} $ and the partition function 
$ Z={\rm tr} \left(e^{-H_{{\rm ent}}}\right) $. 
Furthermore, the average $ \langle \mathcal{O} \rangle $
of a local operator in subsystem A can be calculated as $ \langle \mathcal{O} \rangle = {\rm tr}(\rho_{A} \mathcal{O}_{A}) $.

By tracing out a single spin direction, e.g., the negative spin $ \downarrow $, 
from the ground state on the honeycomb lattice in the presence of the s-wave and chiral $ d+id $-wave superconductivity, the correlation matrix can be 
formulated as
\begin{equation}
C(\vec k)=
\left( 
\begin{array}{cc}
\langle a_{\vec{k}\uparrow}^{\dagger}a_{\vec{k}\uparrow} \rangle & 
\langle a_{\vec{k}\uparrow}^{\dagger}b_{\vec{k}\uparrow} \rangle \\
\langle b_{\vec{k}\uparrow}^{\dagger}a_{\vec{k}\uparrow} \rangle &
\langle b_{\vec{k}\uparrow}^{\dagger}b_{\vec{k}\uparrow} \rangle
\end{array} \right).
\label{corr}
\end{equation}
For more technical details of the analytical calculations of the correlation
matrix, we refer the reader to Appendix (\ref{app2}).
Here, one can show that the 
eigenvalues of the correlation matrix $ \eta_{l} $ 
are related to the entanglement spectrum $ \xi_{l} $, such that
\begin{align}
\xi_{l}=\ln\left(\frac{1-\eta_{l}}{\eta_{l}}\right). \label{ent_levels}
\end{align}

\subsection{s-wave scenario}
\begin{figure}[t]
\includegraphics[width=0.7\columnwidth]{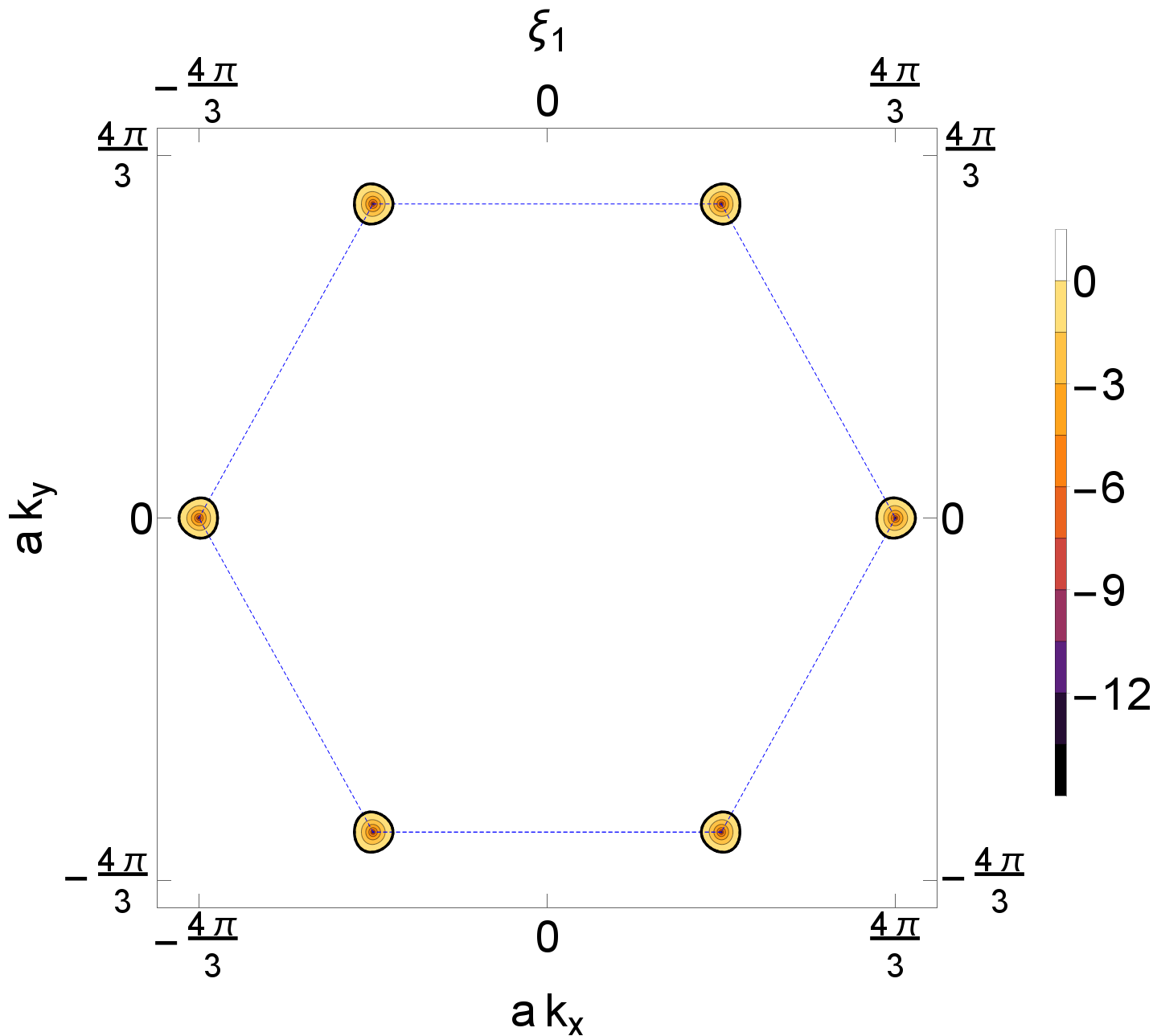}
\includegraphics[width=0.7\columnwidth]{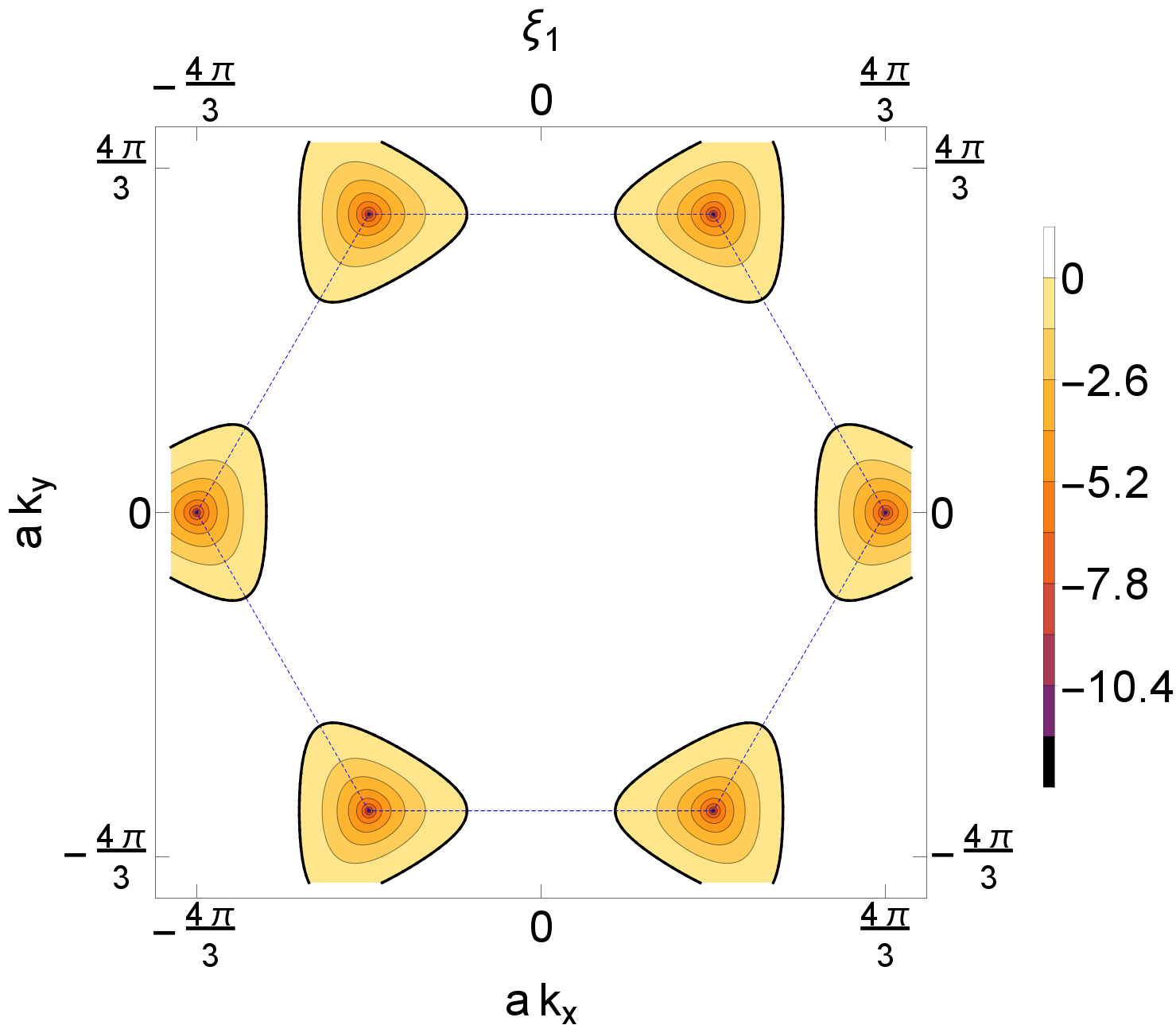}
\includegraphics[width=0.7\columnwidth]{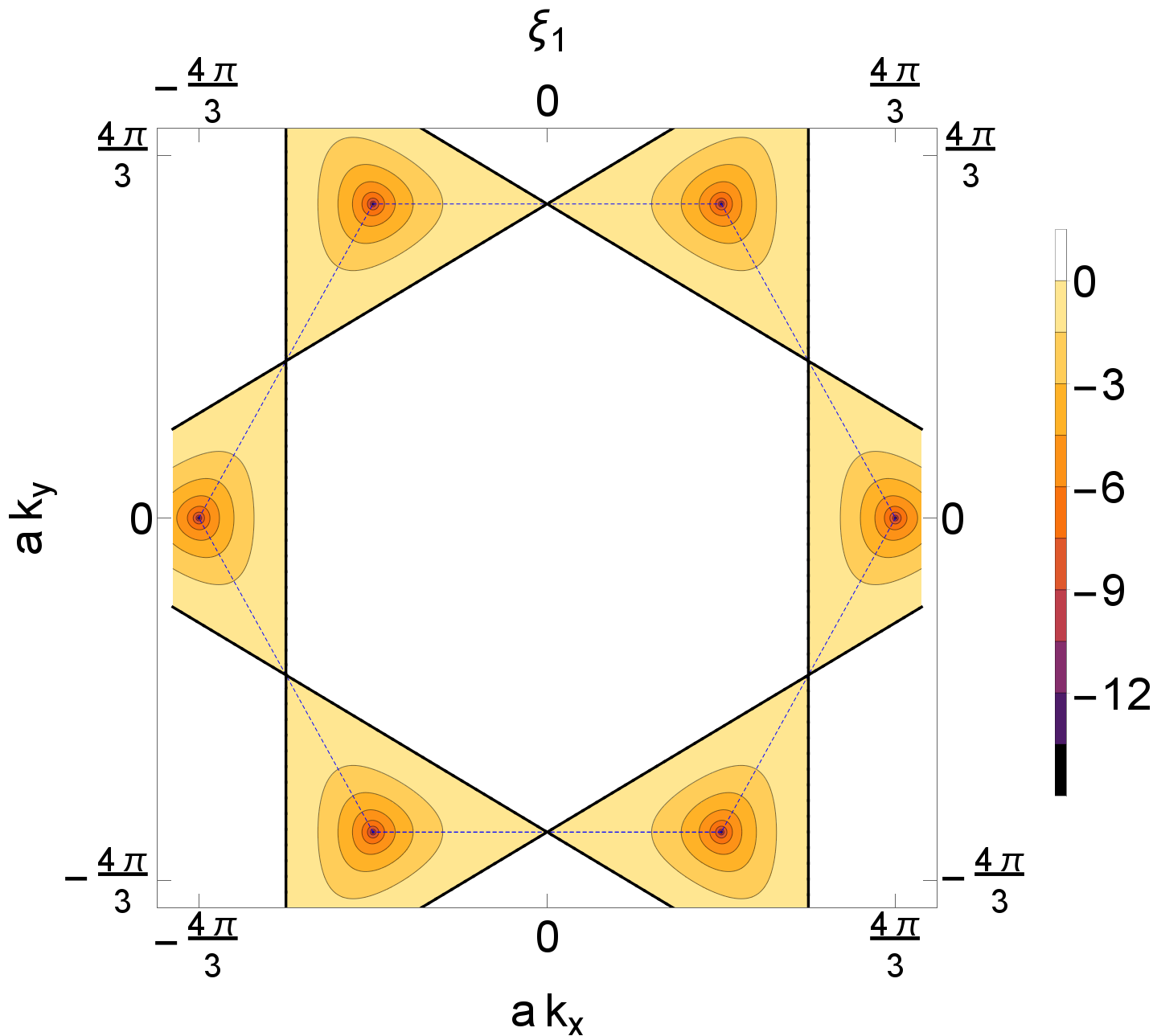}
\caption{(Color online)
Contour plot of entanglement level $  \xi_{1}(\vec{k}) $ of s-wave superconductivity
state on honeycomb lattice plotted for $ \frac{J}{t}=3 $ and:
 (a) $ \frac{\mu}{t}=0.2 $; (b)  $ \frac{\mu}{t}=0.8 $; and 
(c) $ \frac{\mu}{t}=1 $.
The thin blue dashed and thick black lines represent the first Brillouin zone and connect the zero energy states, respectively.
}
\label{fig2}
\end{figure}

The s-wave superconductivity order parameter corresponds to the bond-independent 
superconductivity state; thus, $ S_{\vec{k}} $ is identically zero. 

We analytically obtain the entanglement levels (Eq.~(\ref{ent_levels}))
\begin{eqnarray}
\xi_{1}(\vec{k})=-2{\rm arcsinh}\left(\frac{t \vert \gamma(\vec{k})\vert +\mu}{\vert C_{\vec{k}}\vert}\right) \label{ents1}
\end{eqnarray}
and
\begin{eqnarray}
\xi_{2}(\vec{k})=2{\rm arcsinh}\left(\frac{t \vert \gamma(\vec{k})\vert -\mu}{\vert C_{\vec{k}}\vert}\right)
\label{ents2}.
\end{eqnarray}
The entanglement Hamiltonian has the form
\begin{equation}
\mathcal{H}_{ent}=\sum_{\vec{k}}\left(\xi_{1}e_{\vec{k},+}^{\dagger}
e_{\vec{k},+}
+\xi_{2}f_{\vec{k},+}^{\dagger}f_{\vec{k},+}\right),
\end{equation}
where $ e_{\vec{k},+} $ and $ f_{\vec{k},+} $ are Bogoliubov 
transformations given in Appendix \ref{app2} by Eq.(\ref{BogoliubovSCs1},
\ref{BogoliubovSCs2}).
The entanglement levels for different 
values of $ \mu $, with $ t=2.5eV $, and $ \Delta=3 eV $ are shown in Fig. \ref{fig2}.

The undoped graphene is a gapless semi-metal and is not a superconductor at 
low temperatures.
However, when the system is at half-filling (with $ \mu = 0 $),
the entanglement levels are
\begin{eqnarray}
\xi_{1,2}(\vec{k})=\pm 2{\rm arcsinh}\left(\frac{t}{\Delta}\right),
\end{eqnarray}  
being constant over the entire Brillouin zone. 
In the strong coupling regime, when $ \Delta\gg t $, one finds
\begin{eqnarray}
\xi_{1,2}(\vec{k})\approx \pm  2\frac{t}{\Delta}.
\end{eqnarray} 
The canonical entanglement Hamiltonian at half-filling 
is independent of the inverse
temperature \cite{Schliemann14} $ \beta=k_{E}/\Delta $, such that
\begin{equation}
\mathcal{H}_{can}=\sum_{i=1}^{2}\frac{1}{k_{E}}\left(e_{\vec{k},+}^{\dagger}
e_{\vec{k},+}+f_{\vec{k},+}^{\dagger}f_{\vec{k},+}\right),
\end{equation}
where $ k_{E} $ is a constant.
In general, there is no proportionality between the entanglement Hamiltonian 
and the energy Hamiltonian of free fermions, because the coupling between subsystems 
$ C_{\vec{k}} $  is $ \vec{k} $-dependent in the Brillouin zone. 
When $ C_{\vec{k}}=0 $, at the Dirac points,
the entanglement levels are not entangled. However, at finite doping, the maximally entangled
states, when the entanglement levels are zero, correspond to the zero energy state
of the noninteracting fermions. 
To provide a superior visualization, a thick black line is used to connect the zero-energy states in Fig. \ref{fig1} and the maximally entangled states in Fig. \ref{fig2}. 

\subsection{chiral d-wave scenario}

To enable analytical calculations, we diagonalize the Hamiltonian (\ref{HamS})
\begin{align}
H_{MF}=&\sum_{\vec{k}}E_{\alpha}(o_{\vec{k},+}^{\dagger}
o_{\vec{k},+} + o_{-\vec{k},-}^{\dagger}o_{-\vec{k},-}) \nonumber \\
+ &\sum_{\vec{k}}E_{\beta}(p_{\vec{k},+}^{\dagger}
p_{\vec{k},+} + p_{-\vec{k},-}^{\dagger}p_{-\vec{k},-})
\end{align}
by the Bogoliubov quasiparticles $ o_{\vec{k},+}, o_{-\vec{k},-}, p_{\vec{k},+} $ and 
$ p_{-\vec{k},-} $ given in the Appendix (\ref{app1}) with Eqs.(\ref{Bogoliubov_o_d})-(\ref{Bogoliubov_p_f}).
The energies of Bogoliubov quasiparticles are
$ \pm E_{\alpha} $ and $ \pm E_{\beta} $, where
\begin{equation}
E_{\alpha}=\sqrt{t^{2}\vert \gamma(\vec{k})\vert^{2}+\mu^{2}+\left(\vert S_{\vec{k}}\vert^{2}+\vert C_{\vec{k}}\vert^{2}\right)+ 2\sqrt{u+v}} \label{e_alpha_d}
\end{equation}
and
\begin{equation}
E_{\beta}=\sqrt{t^{2}\vert \gamma(\vec{k})\vert^{2}+\mu^{2}+\left(\vert S_{\vec{k}}\vert^{2}+\vert C_{\vec{k}}\vert^{2}\right)- 2\sqrt{u+v}} \label{e_beta_d}
\end{equation}
with
\begin{align}
u=\left(\mu^{2}+\vert S_{\vec{k}}\vert^{2}\right)t^{2}\vert \gamma(\vec{k})\vert^{2}, \label{def_u}
\end{align} and
\begin{align}
v=\left({\rm Re}(C_{\vec{k}}){\rm Im}(S_{\vec{k}})-{\rm Re}(S_{\vec{k}}){\rm Im}
(C_{\vec{k}})\right)^{2} \label{def_v}.
\end{align} 
When the superconductivity order parameters $ \Delta_{\vec{\delta}} $ are pure
real, i.e., when no time-reversal symmetry breaking occurs, $ v $ vanishes.

From analytical calculations, one obtains the correlation matrix at $ T=0 $
\begin{align}
C(\vec k)=&\left(
\begin{array}{cc}
  C_{11}(\vec{k}) & C_{12}(\vec{k}) \\
 C_{12}^{\ast}(\vec{k}) & C_{22}(\vec{k})\\
\end{array}\right) ,
\label{corrd1}
\end{align}
where
\begin{small}
\begin{align}
&C_{11}=\langle a_{\vec{k}\uparrow}^{\dagger} a_{\vec{k}\uparrow} \rangle \nonumber \\
=&\frac{1}{2}
+\frac{1}{4}\frac{\mu}{\sqrt{\mu^{2}+\vert S_{\vec{k}}\vert^{2}}}(\epsilon_{1}+m)\frac{1}{E_{\alpha}}\left(1-\frac{m}{\sqrt{t^{2}\vert \gamma(\vec{k})\vert^{2}+m^{2}}}\right) \nonumber \\
&+\frac{1}{4}\frac{\mu}{\sqrt{\mu^{2}+\vert S_{\vec{k}}\vert^{2}}}(\epsilon_{2}+m)\frac{1}{E_{\beta}}\left(1+\frac{m}{\sqrt{t^{2}\vert \gamma(\vec{k})\vert^{2}+m^{2}}}\right), \\
&C_{22}=\langle b_{\vec{k}\uparrow}^{\dagger} b_{\vec{k}\uparrow} \rangle \nonumber 
\\=&\frac{1}{2}
+\frac{1}{4}\frac{\mu}{\sqrt{\mu^{2}+\vert S_{\vec{k}}\vert^{2}}}(\epsilon_{1}-m)\frac{1}{E_{\alpha}}\left(1+\frac{m}{\sqrt{t^{2}\vert \gamma(\vec{k})\vert^{2}+m^{2}}}\right) \nonumber \\
&+\frac{1}{4}\frac{\mu}{\sqrt{\mu^{2}+\vert S_{\vec{k}}\vert^{2}}}(\epsilon_{2}-m)\frac{1}{E_{\beta}}\left(1-\frac{m}{\sqrt{t^{2}\vert \gamma(\vec{k})\vert^{2}+m^{2}}}\right), \\ 
&C_{12}=\langle a_{\vec{k}\uparrow}^{\dagger} b_{\vec{k}\uparrow} \rangle \nonumber \\
=&\frac{1}{4}e^{-i\phi_{\vec{k}}}\left(\left(\frac{\epsilon_{1}}{E_{\alpha}}-\frac{\epsilon_{2}}{E_{\beta}}\right)-in\left(\frac{1}{E_{\alpha}}-\frac{1}{E_{\beta}}\right)\right)\frac{t\vert \gamma(\vec{k})\vert}{\sqrt{t^{2}\vert \gamma(\vec{k})\vert^{2}+m^{2}}} \label{elem_corr}
\end{align}
\end{small}
with
\begin{equation}
\epsilon_{1,2}=\sqrt{\mu^{2}+\vert S_{\vec{k}}\vert^{2}}\pm
\sqrt{t^{2}\vert \gamma(\vec{k})\vert^{2}+m^{2}},
\end{equation}
while
\begin{equation}
m=\frac{{\rm Re(C_{\vec{k}})}\cdot{\rm Im(S_{\vec{k}})}-{\rm Im(C_{\vec{k}})}\cdot{\rm Re(S_{\vec{k}})}}{\sqrt{\mu^{2}+ \vert S_{\vec{k}}\vert^{2}}}, \label{h12}
\end{equation}
and
\begin{equation}
n=\frac{{\rm Re}(C_{\vec{k}}){\rm Re}(S_{\vec{k}})+{\rm Im}(C_{\vec{k}}){\rm Im}(S_{\vec{k}})}{\sqrt{\mu^{2}+\vert S_{\vec{k}}\vert^{2}}}.
\end{equation}

Thus, the entanglement spectrum obtained from the eigenvalues of the correlation matrix
given in Eq. (\ref{ent_levels}) consists of entanglement levels 
$ \xi_{1} $ and $ \xi_{2} $ where:
\begin{small}
\begin{align}
\xi_{1,2}=-2 \artanh(C_{11}+C_{22}-1\pm \sqrt{\left(C_{11}-C_{22}
\right)^{2}+4\vert C_{12}\vert^{2}}).
\end{align}
\end{small}

\begin{figure}[t]
\includegraphics[width=0.7\columnwidth]{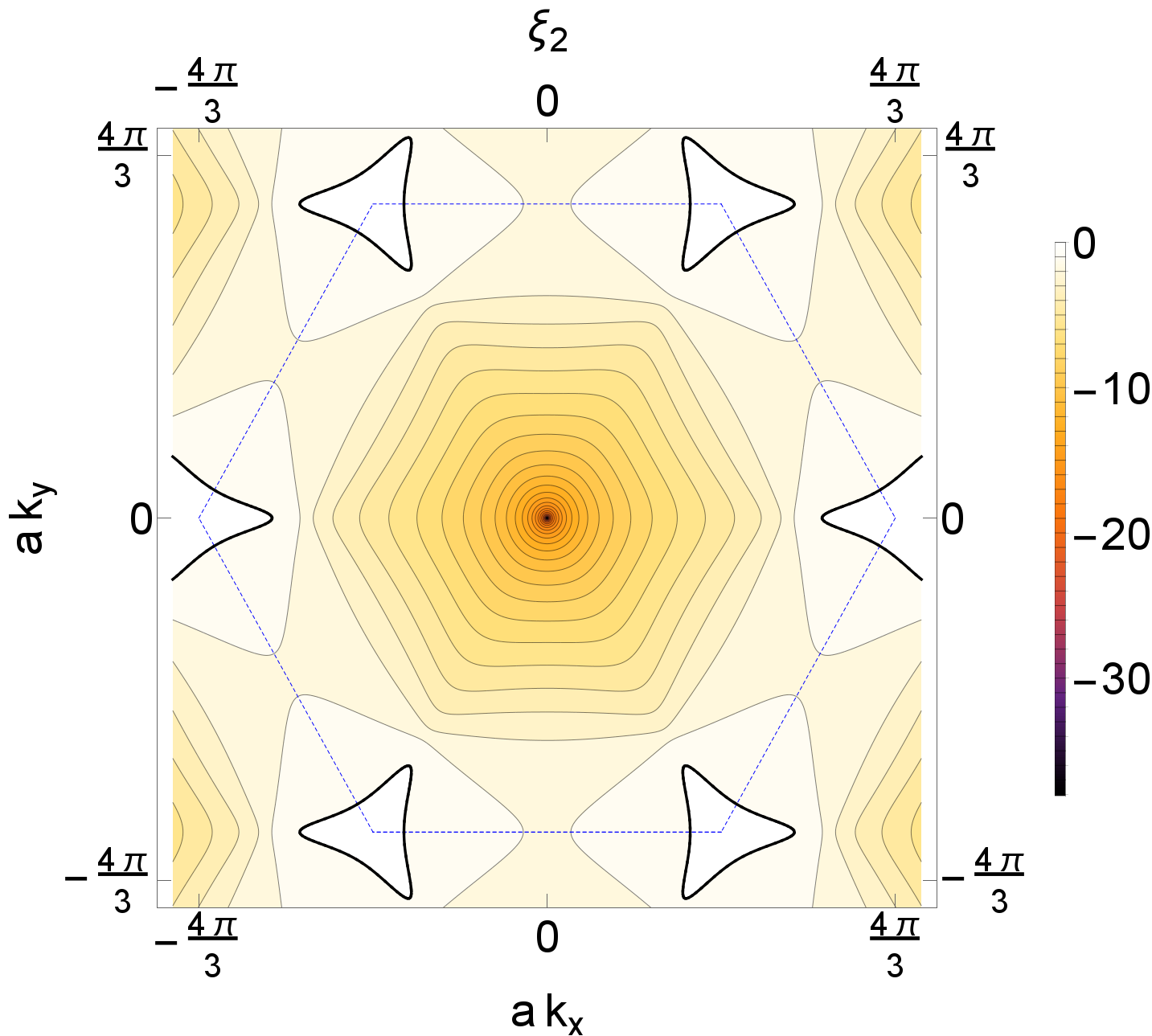}
\includegraphics[width=0.7\columnwidth]{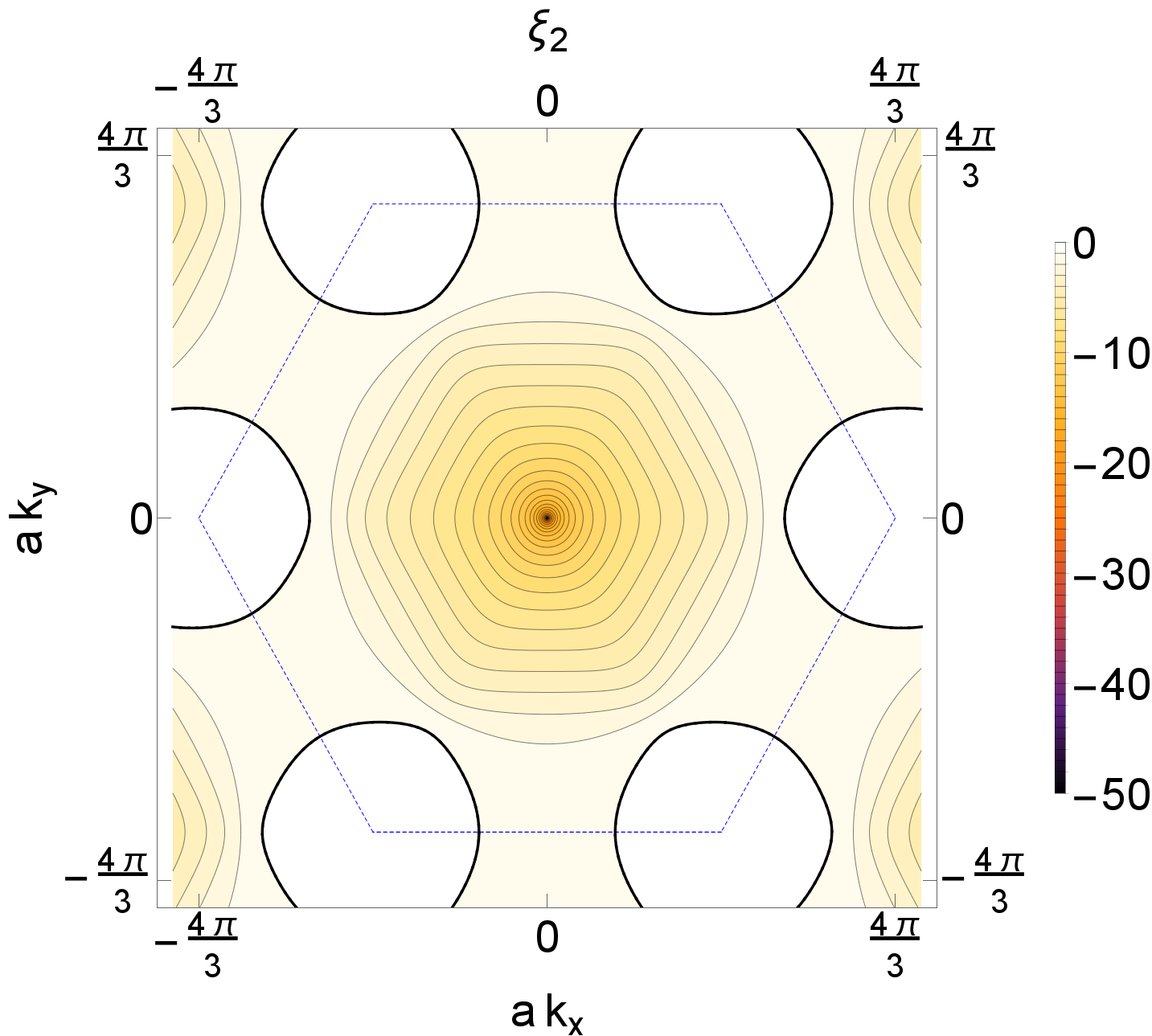}
\includegraphics[width=0.7\columnwidth]{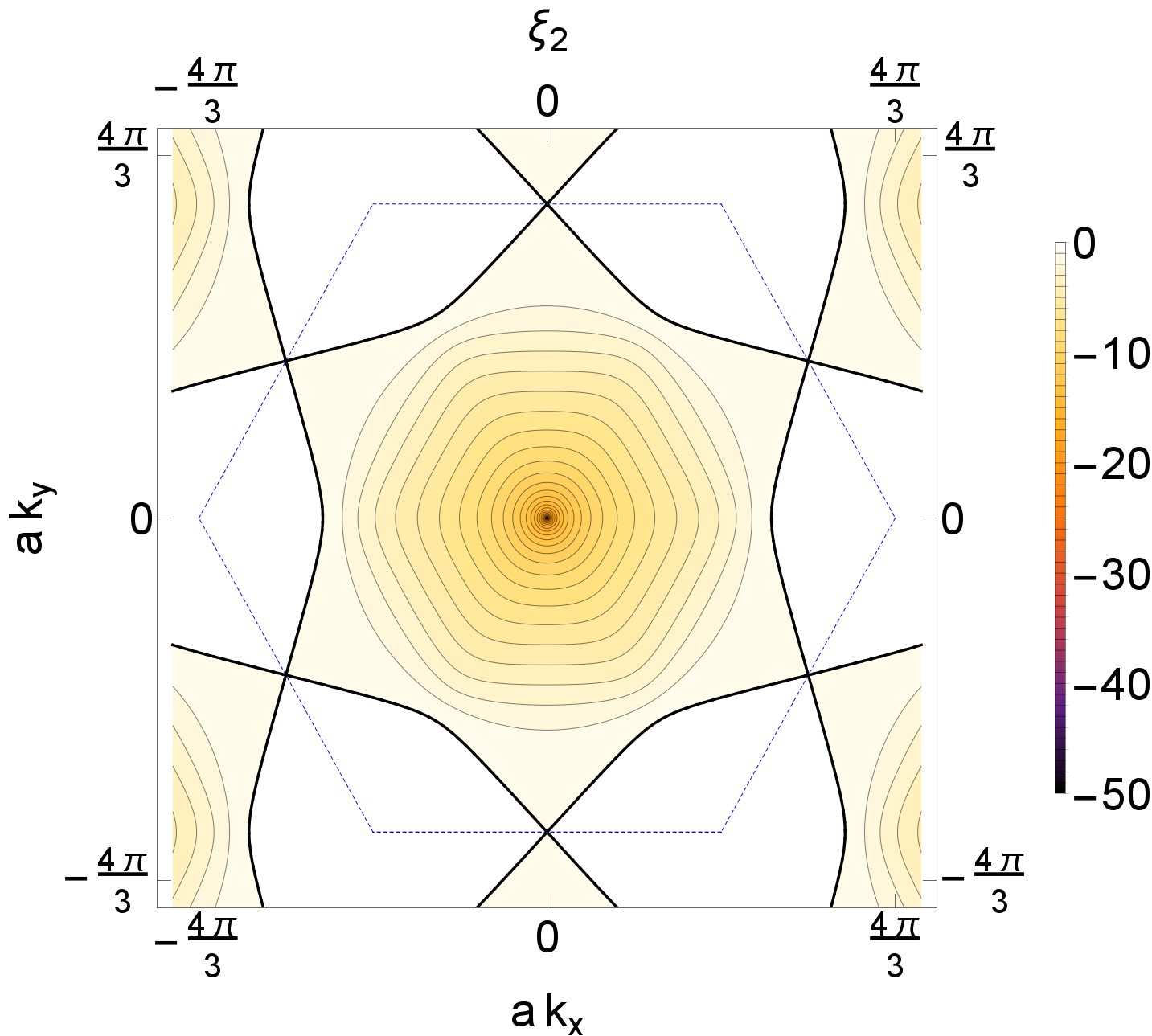}
\caption{(Color Online)
Contour plot of entanglement level $ \xi_{1}(\vec{k}) $  of $ d_{x^{2}-y^{2}}+id_{xy} $
-wave superconductivity state on honeycomb lattice
plotted for $ \frac{J}{t}=3 $ and 
a) $ \frac{\mu}{t}=0.2 $, b)  $ \frac{\mu}{t}=0.8 $ and 
c) $ \frac{\mu}{t}=1 $.
The dashed blue line delineates the first Brillouin zone, while the thick 
black line shows maximally entangled states.
}
\label{fig3}
\end{figure}

As the d-wave spin-singlet superconductivity order
parameter involves both $ C_{\vec{k}} $ and $ S_{\vec{k}} $,
there is no relationship between states with the zero-value states 
of the entanglement spectrum 
and the zero-energy states of the free fermions.
At the van-Hove singularity point, i.e., when $ \mu = t $, 
both the entanglement spectrum and the 
energy spectrum of the free fermions are zero at the $ M $ point. The results
of our analytical calculations
of the entanglement spectrum of the $ d_{x^{2}-y^{2}}+id_{xy} $-wave superconductivity 
on the honeycomb lattice are presented in Fig. \ref{fig3}. 

As we have discussed above, the $ d_{x^{2}-y^{2}} $- and
$ d_{xy} $-wave superconductivity order 
parameters preserve the time-reversal symmetry 
(Eq.~(\ref{time})). Based on the time-reversal symmetry and provided  
$ \Psi_{\vec{k}} $ are the eigenstates of the Hamiltonian given in Eq.(\ref{HamS}),
we can state that
\begin{equation}
\Psi_{\vec{k}}^{\ast}=\Psi_{-\vec{k}}
\end{equation}
where the $ \Psi_{-\vec{k}}^{\ast} $ are also eigenstates of the Hamiltonian
of Eq.~(\ref{HamS}).
This yields
\begin{equation}
\Phi_{\vec{k}}^{\ast}=\Phi_{-\vec{k}}.
\end{equation}
Hence, the real $ d $-wave superconductivity order parameter 
preserves the time-reversal symmetry in the correlation matrix, 
which is constructed from the $ \Phi_{\vec{k}} $ as 
$ C(\vec{k})=\langle \Phi_{\vec{k}}^{\dagger} \Phi_{\vec{k}} \rangle $. 
The entanglement Hamiltonian satisfies:
\begin{align}
T_{E} \mathcal{H}_{ent}(\vec{k}) T_{E}^{-1}=\mathcal{H}_{ent}(-\vec{k}), \label{time1}
\end{align}
with $ T_{E}=K $.

When the $ d_{x^{2}-y^{2}}+id_{xy} $-wave superconductivity order parameter is 
considered, $ C_{\vec{k}} $ and $ S_{\vec{k}} $ are complex functions. Then,
the $ m $ and $ n $ terms are non-zero. Hence,
the average occupancy number at site A, $ C_{11}(\vec{k}) $,
and the average occupancy number at site B, $ C_{22}(\vec{k}) $, 
are inequivalent and the off-diagonal element of the correlation matrix
$ C_{12}(\vec{k}) $ is complex.
Because $ S_{\vec{k}} $ is an odd function in the momentum space, while $ 
C_{\vec{k}} $ is a even function, it can be shown that elements of the correlation
matrix $ C_{11}(\vec{k}) $, $ C_{22}(\vec{k}) $, and $ C_{12}(\vec{k}) $
are constrained as $ C_{11}(-\vec{k})=C_{22}(\vec{k}) $ and 
$ C_{12}^{\ast}(-\vec{k})=C_{12}(\vec{k}) $.
Therefore, it follows that the complex $ d_{x^{2}-y^{2}}+id_{xy} $-wave superconductivity order
parameter breaks the time-reversal symmetry in the entanglement 
Hamiltonian. The topology of the entanglement Hamiltonian in two-dimensions with 
broken time-reversal symmetry is characterized by the entanglement Chern number.

For further analysis of the topological properties of the entanglement Hamiltonian,
we require not only its eigenvalues, but also its eigenstates.
The eigenstates of the correlation matrix are identical to the eigenstates
of the entanglement Hamiltonian and can be expressed as
\begin{align}
q_{\vec{k}\uparrow}=&\delta_{+}(\vec{k})a_{\vec{k}\uparrow}+
\delta_{-}(\vec{k})b_{\vec{k}\uparrow} \\
r_{\vec{k}\uparrow}=&\delta_{+}(-\vec{k})a_{\vec{k}\uparrow}-
\delta_{-}^{\ast}(-\vec{k})b_{\vec{k}\uparrow}
\end{align}
where explicit expressions for $ \delta_{+}(\vec{k}) $ and 
$ \delta_{-}(\vec{k}) $ are given in Appendix (\ref{app2}) by Eq.(\ref{delta}).
Using these eigenstates, we can calculate the Berry curvature
\begin{equation}
  F(\vec k)=\frac{\partial A_y}{\partial k_x}
  -\frac{\partial A_x}{\partial k_y}
\label{berrycurv}
\end{equation}
and the Berry connection
\begin{equation}
  \vec A(\vec k)=
  i\langle r(\vec k)|
  \frac{\partial}{\partial\vec k}|r(\vec k)\rangle,
\label{berrycon}
\end{equation}
which vanish everywhere outside the Dirac points where quantized ”monopole” 
sources of the $ \delta $-function type exist.

Through numerical integrations of the Berry curvature along the Brillouin zone,
we find that the entanglement Chern number is $ C=1 $, in the case
of the chiral $ d_{x^{2}-y^{2}}+id_{xy} $-wave superconductivity state. 
In the presence of $ SU(2) $ rotation and broken time-reversal
symmetry, as in the case of an energetic Hamiltonian, the Chern number $ C $ can have
even values only. For the entanglement Hamiltonian,
it is possible to obtain an odd value for the Chern
number, as it is not invariant to the $ SU(2) $ rotation.
As a result, the topology of the entanglement Hamiltonian, which is 
obtained by tracing out the spin-down subsystem of the ground 
state of the chiral $ d_{x^{2}-y^{2}}+id_{xy} $-wave superconductivity 
state on the honeycomb lattice, clearly differs from the topology of 
the energetic Hamiltonian of free fermions without the superconductivity 
instabilities.

\subsection{tracing out B sublattices}

\subsubsection{s-wave scenario}

\begin{figure}[t]
\includegraphics[width=0.7\columnwidth]{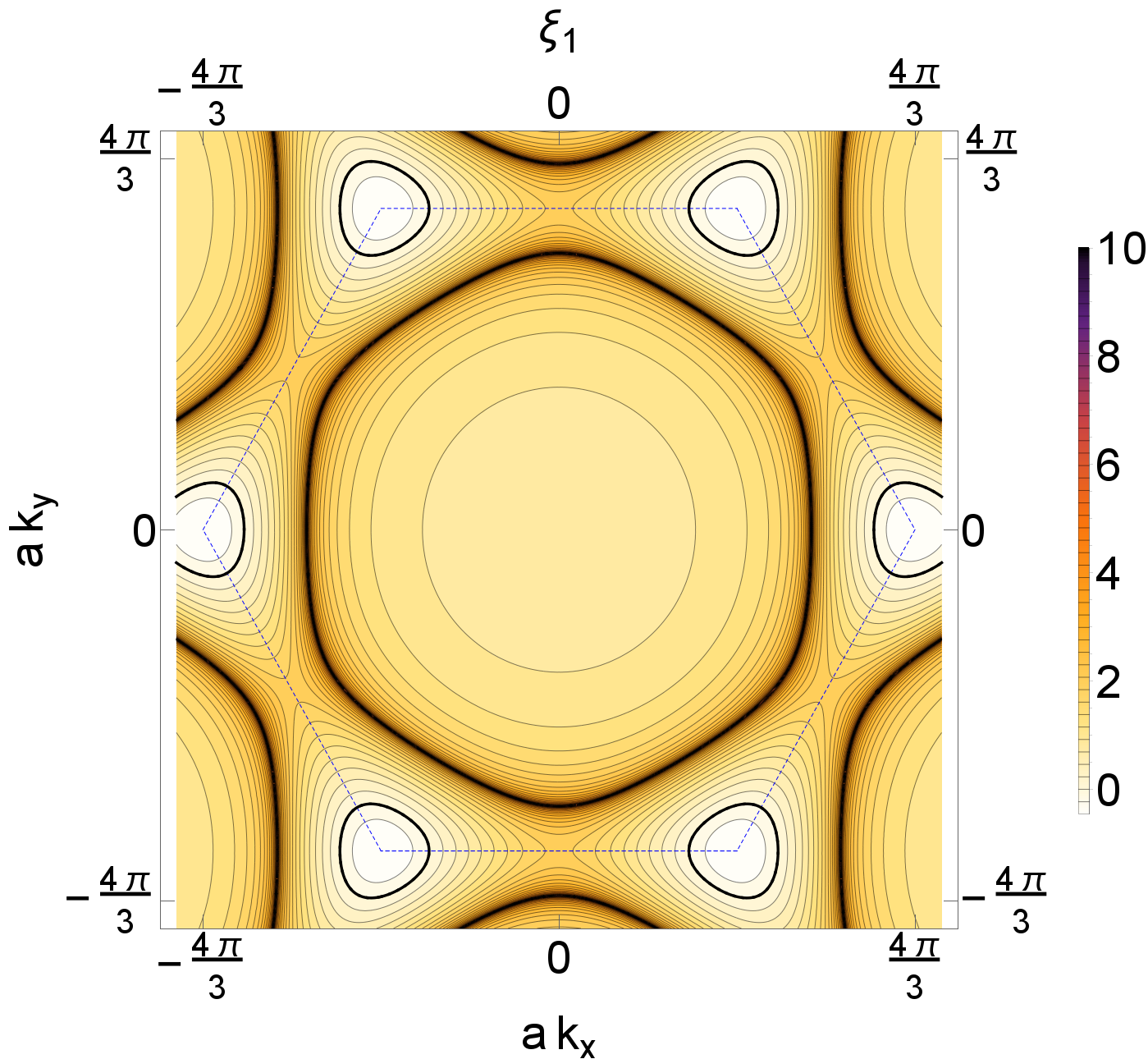}
\caption{(Color online)
Contour plot of entanglement level $  \xi_{1}(\vec{k}) $ of s-wave superconductivity
state on honeycomb lattice plotted for $ \frac{J}{t}=3 $ and
 $ \frac{\mu}{t}=0.8 $.
The first Brillouin zone is border
by the dashed blue line, while the thick line connects 
maximally entangled states.
}
\label{fig4}
\end{figure}

We will now consider the ground state of interacting fermions 
on the honeycomb lattice in the presence of the s-wave 
superconductivity instability. Upon tracing out B sublattices the entanglement levels:
\begin{align}
\xi_{\pm} = \pm 2 {\rm arctanh}\left(\frac{t^{2}\vert \gamma (\vec{k})\vert^{2}-\mu^{2}+\Delta^{2}\vert\gamma(\vec{k})\vert^{2}}{E_{\alpha}E_{\beta}}\right)
\end{align}
where $ E_{\alpha}=\sqrt{\left(t\vert\gamma(\vec{k})\vert-\mu\right)^{2}+\Delta^{2}\vert\gamma(\vec{k})\vert^{2}} $
and $ E_{\beta}=\sqrt{\left(t\vert\gamma(\vec{k})\vert+\mu\right)^{2}+\Delta^{2}\vert\gamma(\vec{k})\vert^{2}}$.
When system is at half-filling the subsystems are maximally entangled.
The entanglement levels are plotted at Fig. \ref{fig4}.

\subsubsection{chiral d-wave scenario}

\begin{figure}[t]
\includegraphics[width=0.7\columnwidth]{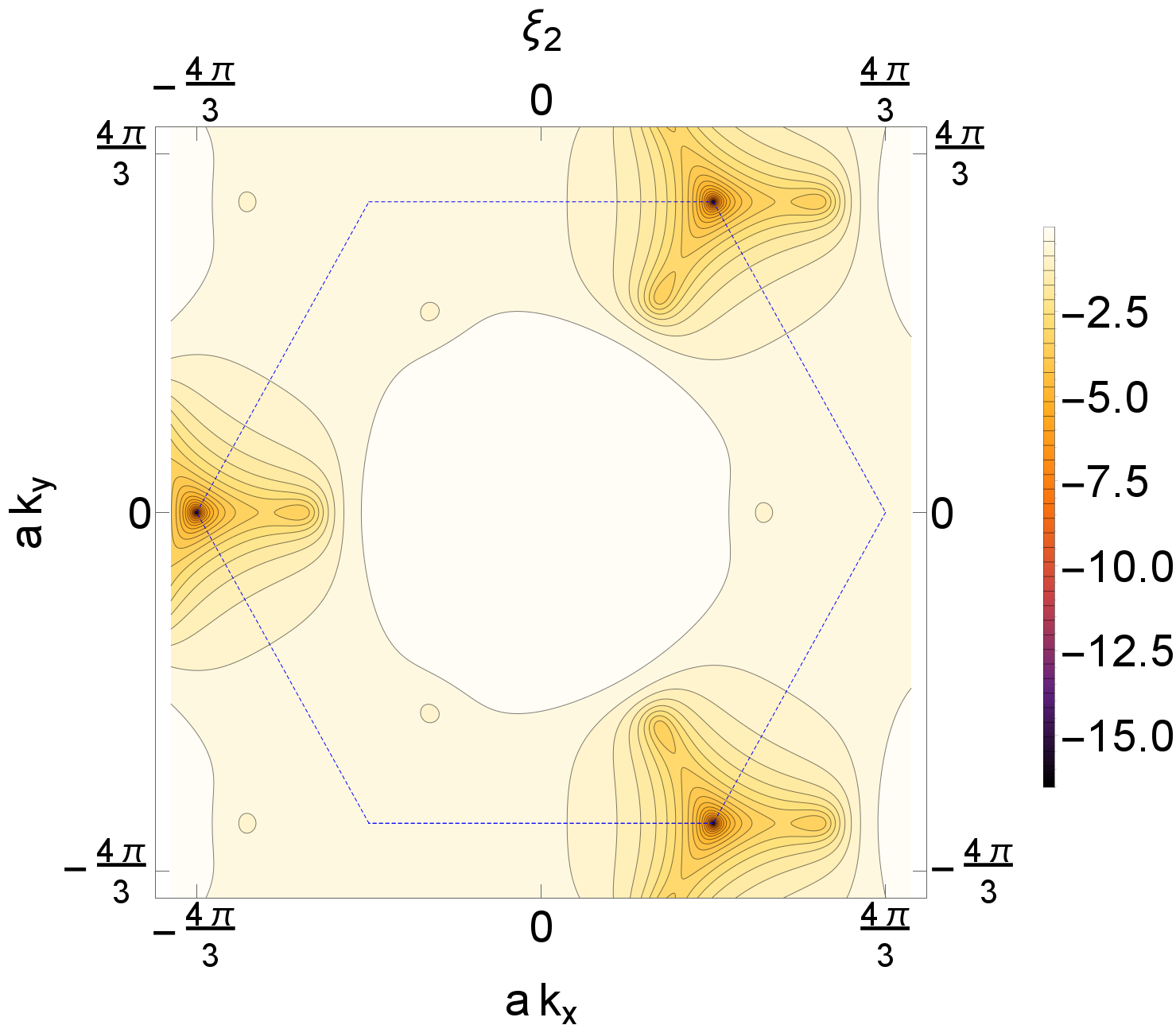}
\caption{(Color online)
Contour plot of entanglement level $  \xi_{2}(\vec{k}) $ of $ d_{x^{2}-y^{2}}+id_{xy} $ superconductivity
state on honeycomb lattice plotted for $ \frac{J}{t}=3 $ and $ \frac{\mu}{t}=0.8 $.
The thin blue dashed and thick black lines represent the first Brillouin zone and connect the zero energy states, respectively.
}
\label{fig5}
\end{figure}

Upon tracing out B sublattices, the entanglement spectrum
of d-wave superconductivity state on the honeycomb lattice is
completely determined by the correlation matrix:
\begin{align}
C(\vec{k})=\left(\begin{array}{cc}
C_{11}(\vec{k}) & C_{13}(\vec{k}) \\
C_{13}^{\ast}(\vec{k}) & C_{33}(\vec{k})
\end{array}\right)
\end{align}
where $ C_{11} $, $ C_{33} $ and $ C_{13} $ are given in the Appendix (\ref{app2}). The eigenvalues $ \eta_{1,2} $ of the correlation matrix 
\begin{small}
\begin{align}
\eta_{1,2}=\frac{1}{2}\left((C_{11}+C_{33}+\pm \sqrt{\left(C_{11}-C_{33}
\right)^{2}+4\vert C_{13}\vert^{2}}\right).
\end{align}
\end{small}
are related to the entanglement levels $ \xi_{1,2}={\rm ln}\left(\frac{\eta_{\pm}}{1-\eta_{pm}}\right) $. At finite doping
the entanglement levels never vanish. Here, space inversion symmetry
of the entanglement spectrum is broken and the entanglement levels
satisfy $ \xi_{\pm}(-\vec{k})=-\xi_{\mp}(\vec{k}) $. The entanglement level $ \xi_{2}(\vec{k}) $ is visualized in
Fig \ref{fig5}. The broken time-reversal symmetry in the entanglement Hamiltonian
leads to the entanglement Chern number $ C=1 $.

\section{Conclusion and Outlook}
\label{conclusion}
We analytically evaluated the entanglement spectra of the superconductivity 
states on the graphene honeycomb lattice, 
primarily focusing on the s-wave and chiral 
$ d_{x^{2}-y^{2}}+id_{xy} $ superconductivity states. 
When one spin direction was traced out,
exact correspondence between the maximally entangled states of the s-wave
superconductor and the zero energies of the noninteracting fermionic honeycomb
lattice at finite doping was observed. The relationship between the topologies 
of the entanglement and subsystem Hamiltonians was found to depend on the coupling 
between the subsystems. 
Further, the chiral $ d_{x^{2}-y^{2}}+id_{xy} $ superconductivity
order parameter breaks the time-reversal symmetry in the entanglement Hamiltonian. 
The topological properties of the entanglement Hamiltonian, 
characterized by the topological nontrivial entanglement Chern number $ C = 1 $, 
clearly differ from those of the time-reversal invariant Hamiltonian of 
the noninteracting fermions on the honeycomb lattice. The
investigations presented herein are based on closed analytical expressions for the full 
eigensystems of the s- and d-wave superconductivity states on the honeycomb lattice over 
the entire Brillouin zone. The method used to examine these eigensystems
may constitute a useful tool for new studies of superconductivity in graphene. 
Future work may investigate the relationship between the topologies
of the entanglement and subsystem Hamiltonians through the topological phase transition;
for example, in the coexistence region between antiferromagnetism and 
$ d_{x^{2}-y^{2}} +i d_{xy} $
superconducting correlations
in graphene \cite{Black-Schaffer15} and graphene bilayers 
\cite{Milovanovic12}.

\section*{ACKNOWLEDGMENTS}

The authors kindly acknowledge Milica V. Milovanovi\'c.
This work was supported by Deutsche Forschungsgemeinschaft via GRK1570.

\newpage
\appendix
\begin{widetext}
\section{Derivation of the eigensystem } \label{app1}
In this Appendix we present analytical diagolazation of the Hamiltonian 
of the chiral $ d+id $-wave superconductivity state on the honeycomb lattice. 
Complexity of the order parameter makes the analytical approach more difficult.
The starting point of our analysis is the Bardeen-Cooper-Schrieffer 
mean-field Hamiltonian in momentum space is
\begin{align}
H_{MF}(\vec{k})=&-t\sum_{\vec{k}}\left(\gamma(\vec{k})a_{\vec{k}\sigma}^{\dagger}
b_{\vec{k}\sigma}+h.c.\right) 
-\mu\sum_{\vec{k}}\left(a_{\vec{k}\sigma}^{\dagger}a_{\vec{k}\sigma} 
+b_{\vec{k}\sigma}^{\dagger}b_{\vec{k}\sigma}\right)
-J \sum_{\vec{k},\vec{\delta}}\left(\Delta_{\vec{\delta}}e^{i\vec{k}\vec{\delta}}
\left(a_{\vec{k}\uparrow}^{\dagger}b_{-\vec{k}\downarrow}^{\dagger}-
a_{\vec{k}\downarrow}^{\dagger}b_{-\vec{k}\uparrow}
^{\dagger}\right)+h.c.\right) \label{HamMF2}
\end{align} 
where we define the superconductivity order parameter
\begin{equation}
\Delta(\vec{k})=\sum_{\vec{\delta}}\Delta_{\vec{\delta}}e^{i\vec{k}\vec{\delta}}
\end{equation}
as a combination of the $ d_{x^{2}-y^{2}} $ and $ d_{xy} $-wave superconductivity 
state $ \Delta_{d\pm id}(\vec{k})=\cos\left(\frac{\pi}{3}\right)\Delta_{d_{x}^{2}-y^{2}}
(\vec{k})\pm \sin\left(\frac{\pi}{3}\right)\Delta_{d_{xy}}(\vec{k}) $
which minimalizes a free energy. 

We apply the transformations
\begin{eqnarray}
c_{\vec{k},\sigma}=\frac{1}{\sqrt{2}}(a_{\vec{k},\sigma}-e^{i \cdot \phi_{\vec{k}}}b_{\vec{k},\sigma}) 
,\nonumber \\
 d_{\vec{k},\sigma}=\frac{1}{\sqrt{2}}(a_{\vec{k},\sigma}+e^{i \cdot \phi_{\vec{k}}}b_{\vec{k},\sigma})
\end{eqnarray} 
such that in
\begin{equation}
H_{1}(\vec{k})=\left(
\begin{array}{cccc}
t \vert \gamma (\vec{k}) \vert -\mu & 0 & C_{\vec{k}} & 
-i S_{\vec{k}} \\
0 & - t \vert \gamma (\vec{k}) \vert -\mu & i S_{\vec{k}} & - C_{\vec{k}} \\ C_{\vec{k}}^{\ast} & -i S_{\vec{k}}^{\ast} &
- t \vert \gamma (\vec{k}) \vert + \mu & 0\\
i S_{\vec{k}}^{\ast} & - C_{\vec{k}}^{\ast} & 0 & t \vert \gamma (\vec{k}) \vert +\mu
\end{array}
\right). 
\end{equation} 
diagonalize the kinetic part of the Hamiltonian. 
$ C_{\vec{k}} = J \sum_{\vec{\delta}} \vec{\Delta}_{\vec{\delta}}\cos(\vec{k}\vec{\delta} - 
\phi_{\vec{k}}) $ 
and 
$ S_{\vec{k}} = J \sum_{\vec{\delta}} \vec{\Delta}_{\vec{\delta}}\sin(\vec{k}\vec{\delta} - 
\phi_{\vec{k}}) $ are complex functions.

Here it is useful to split this Hamiltonian as $ H_{1}=H_{1}^{'}+H_{1}^{''} $ where
\begin{equation}
H_{1}^{'}(\vec{k})=\left(
\begin{array}{cccc}
t \vert \gamma (\vec{k}) \vert -\mu & 0 & 0 & 
-i S_{\vec{k}} \\
0 & - t \vert \gamma (\vec{k}) \vert -\mu & i S_{\vec{k}} & 0 \\
0 & -i S_{\vec{k}}^{\ast} &
- t \vert \gamma (\vec{k}) \vert + \mu & 0\\
i S_{\vec{k}}^{\ast} & 0 & 0 & t \vert \gamma (\vec{k}) \vert +\mu
\end{array}
\right). 
\end{equation} 
and
\begin{equation}
H_{1}^{''}(\vec{k})=\left(
\begin{array}{cccc}
0 & 0 & C_{\vec{k}} & 0 \\
0 & 0 & 0 & -C_{\vec{k}} \\
C_{\vec{k}}^{\ast} & 0 & 0 & 0\\
0 & -C_{\vec{k}}^{\ast} & 0 & 0
\end{array}
\right).  
\end{equation}
$H_{1}^{'}$ is diagonalized by
\begin{align}
e_{\vec{k}+}=&i\alpha_{-}^{\ast}c_{\vec{k}\uparrow}+\alpha_{+}d_{-\vec{k}\downarrow}
^{\dagger} \\
f_{\vec{k}+}=&-i\alpha_{-}^{\ast}d_{\vec{k}\uparrow}+\alpha_{+}c_{-\vec{k}\downarrow}^{\dagger}
\end{align}
with 
\begin{equation}
\alpha_{+}=\sqrt{\frac{1}{2}\left(1+\frac{\mu}{\sqrt{\mu^{2}+
\vert S_{\vec{k}}\vert^{2}}}\right)} \quad, \quad 
\alpha_{-}=\frac{ S_{\vec{k}}}{\sqrt{2\sqrt{\mu^{2}+\vert S_{\vec{k}}\vert^{2}}\left(\mu+\sqrt{\mu^{2}+\vert S_{\vec{k}}\vert^{2}}\right)}}.
\end{equation}
This leads to
\begin{equation}
H_{2}=U_{2}H_{1}U_{2}^{\dagger}=\left(
\begin{array}{cccc}
e_{1} & m & -l & 0 \\
m & e_{2} & 0 & l \\
-l^{\ast} & 0 & -e_{1} & m \\
0 & l^{\ast} & m & -e_{2}
\end{array}
\right)
\end{equation}
with 
\begin{equation}
m=\frac{{\rm Re(C_{\vec{k}})}\cdot{\rm Im(S_{\vec{k}})}-{\rm Im(C_{\vec{k}})}
\cdot{\rm Re(S_{\vec{k}})}}{\sqrt{\mu^{2}+ \vert S_{\vec{k}}\vert^{2}}}
\end{equation}
and
\begin{equation}
l=\alpha_{+}^{2}C_{\vec{k}}^{\ast}+(\alpha_{-}^{\ast})^{2}C_{\vec{k}}
\end{equation}
and $ \pm e_{1} $ and $ \pm e_{2} $ are eigenenergies of the Hamiltonian $ H_{1}^{'} $
given by
\begin{equation}
e_{1}=t \vert \gamma (\vec{k}) \vert +\sqrt{\mu^{2}+ \vert S_{\vec{k}}\vert^{2}}
\end{equation}
and
\begin{equation}
e_{2}=-t \vert \gamma (\vec{k}) \vert +\sqrt{\mu^{2}+\vert S_{\vec{k}}\vert^{2}}.
\end{equation}
We can now split this Hamiltonian as $ H_{2}=H_{2}^{'}+H_{2}^{''} $ where
\begin{equation}
H_{2}^{'}=\left(
\begin{array}{cccc}
e_{1} & m & 0 & 0 \\
m & e_{2} & 0 & 0 \\
0 & 0 & -e_{1} & m \\
0 & 0 & m & -e_{2}
\end{array}
\right)
\quad,\quad
H_{2}^{''}=\left(
\begin{array}{cccc}
0 & 0 & -l & 0 \\
0 & 0 & 0 & l \\
-l^{\ast} & 0 & 0 & 0 \\
0 & l^{\ast} & 0 & 0
\end{array}
\right) \,.
\end{equation}
Proceeding now with the transformations
\begin{align}
g_{\vec{k}+}=&\beta_{+}e_{\vec{k}+}+\sigma\beta_{-}f_{\vec{k}+} \\
h_{\vec{k}+}=&\sigma\beta_{-}e_{\vec{k}+}-\beta_{+}f_{\vec{k}+}
\end{align}
where $ \sigma={\rm sign}(m) $ and
\begin{equation}
\beta_{\pm}=\sqrt{\frac{1}{2}\left(1\pm\frac{t\vert \gamma(\vec{k})\vert}
{\sqrt{t^{2}\vert \gamma(\vec{k})\vert^{2}+m^{2}}}\right)}
\end{equation}
we diagonalize first part of the Hamiltonian $ H_{2}^{'} $ and we get
\begin{equation}
H_{3}=U_{3}H_{2}U_{3}^{\dagger}=\left(
\begin{array}{cccc}
\epsilon_{1} & 0 & 0 & -l \\
0 & \epsilon_{2} & -l & 0\\
0 & -l^{\ast} & -\epsilon_{2} & 0\\
-l^{\ast} & 0 & 0 & -\epsilon_{1}
\end{array}
\right)
\end{equation}
where $ \pm \epsilon_{1} $ and $ \pm \epsilon_{2} $ are eigenenergies of the Hamiltonian $ H_{2}^{'} $
\begin{equation}
\epsilon_{1}=\sqrt{\mu^{2}+\vert S_{\vec{k}}\vert^{2}}+
\sqrt{t^{2}\vert \gamma(\vec{k})\vert^{2}+m^{2}}
\end{equation}
and
\begin{equation}
\epsilon_{2}=\sqrt{\mu^{2}+\vert S_{\vec{k}}\vert^{2}}-
\sqrt{t^{2}\vert \gamma(\vec{k})\vert^{2}+m^{2}}.
\end{equation}
Finally, this Hamiltonian is brought to the diagonalized form with transformations
\begin{align}
o_{\vec{k}+}=&\gamma_{+}^{(1)}g_{\vec{k}+}-\gamma_{-}^{(1)}g_{\vec{k}-}^{\dagger} 
\label{o}\\
p_{\vec{k}+}=&\gamma_{+}^{(2)}h_{\vec{k}+}-\gamma_{-}^{(2)}h_{\vec{k}-}^{\dagger}
\label{p}
\end{align}
with
\begin{equation}
\gamma_{+}^{(1)}=\sqrt{\frac{1}{2}\left(1+\frac{\epsilon_1}{E_\alpha}\right)}
\quad,\quad 
\gamma_{-}^{(1)}=\frac{l}{\sqrt{2E_{\alpha}\left(E_{\alpha}+\epsilon_{1}\right)}}
\label{defgamma}
\end{equation}
and
\begin{equation}
\gamma_{+}^{(2)}=\sqrt{\frac{1}{2}\left(1+\frac{\epsilon_2}{E_\beta}\right)}
\quad,\quad 
\gamma_{-}^{(2)}=\frac{l}{\sqrt{2E_{\beta}\left(E_{\beta}+\epsilon_{2}\right)}}
\label{defgamma}
\end{equation}
and
\begin{equation}
E_{\alpha}=\sqrt{t^{2}\vert \gamma(\vec{k})\vert^{2}+\mu^{2}+\vert S_{\vec{k}}\vert^{2}+\vert C_{\vec{k}}\vert^{2}+ 2\sqrt{\left(\mu^{2}+\vert S_{\vec{k}}\vert^{2}\right)t^{2}\vert \gamma(\vec{k})\vert^{2}+\left({\rm Re}C_{\vec{k}}{\rm Im}S_{\vec{k}}-{\rm Re}S_{\vec{k}}{\rm Im}C_{\vec{k}}\right)^{2}}} 
\end{equation}
and
\begin{equation}
E_{\beta}=\sqrt{t^{2}\vert \gamma(\vec{k})\vert^{2}+\mu^{2}+\vert S_{\vec{k}}\vert^{2}+
\vert C_{\vec{k}}\vert^{2}- 2\sqrt{\left(\mu^{2}+\vert S_{\vec{k}}\vert^{2}\right)t^{2}
\vert \gamma(\vec{k})\vert^{2}+\left({\rm Re}C_{\vec{k}}{\rm Im}S_{\vec{k}}-{\rm Re}
S_{\vec{k}}{\rm Im}C_{\vec{k}}\right)^{2}}}.
\end{equation}
Bogoliubov transformations $ o_{\vec{k}+} $ and $ p_{\vec{k}+} $ in the basis 
$ a_{\vec{k}\uparrow}, b_{\vec{k},\uparrow} $
\begin{align}
o_{\vec{k}+}=&-\frac{1}{\sqrt{2}}\left(\alpha_{+}\gamma_{-}^{(1)}-i\alpha_{-}^{\ast}
\gamma_{+}^{(1)}\right)\left(\beta_{+}-\sigma \beta_{-}\right) a_{\vec{k}\uparrow}
-\frac{1}{\sqrt{2}}e^{i\phi_{\vec{k}}}\left(\alpha_{+}
\gamma_{-}^{(1)}+i\alpha_{-}^{\ast}\gamma_{+}^{(1)}\right)
\left(\beta_{+}+\sigma \beta_{-}\right)  b_{\vec{k}\uparrow}  \nonumber \\
&+\frac{1}{\sqrt{2}}\left(\alpha_{+}\gamma_{+}^{(1)}+
i\alpha_{-}\gamma_{-}^{(1)}\right) \left(\beta_{+}+\sigma \beta_{-}\right) 
a_{-\vec{k}\downarrow}^{\dagger}
+\frac{1}{\sqrt{2}}e^{i\phi_{\vec{k}}}\left(\alpha_{+}\gamma_{+}^{(1)}-
i\alpha_{-}\gamma_{-}^{(1)}\right)\left(\beta_{+}-\sigma \beta_{-}\right)
b_{-\vec{k}\downarrow}^{\dagger}
\label{Bogoliubov_o_d} \\
p_{\vec{k}+}=&-\frac{1}{\sqrt{2}}\left(\alpha_{+}\gamma_{-}^{(2)}+
i\alpha_{-}^{\ast}\gamma_{+}^{(2)}\right)\left(\beta_{+}+\sigma \beta_{-}\right) 
a_{\vec{k}\uparrow}
+\frac{1}{\sqrt{2}}e^{i\phi_{\vec{k}}}\left(\alpha_{+}\gamma_{-}^{(2)}-
i\alpha_{-}^{\ast}\gamma_{+}^{(2)}\right)
\left(\beta_{+}-\sigma \beta_{-}\right) b_{\vec{k}\uparrow} \nonumber \\
&\frac{1}{\sqrt{2}}\left(\alpha_{+}\gamma_{+}^{(2)}-
i\alpha_{-}\gamma_{-}^{(2)}\right)\left(\beta_{+}-\sigma \beta_{-}\right) 
a_{-\vec{k}\downarrow}^{\dagger}
-\frac{1}{\sqrt{2}}e^{i\phi_{\vec{k}}}\left(\alpha_{+}\gamma_{+}^{(2)}+
i\alpha_{-}\gamma_{-}^{(2)}\right)\left(\beta_{+}+\sigma \beta_{-}\right)
a_{-\vec{k}\downarrow}^{\dagger}
\label{Bogoliubov_p_f}
\end{align}

\section{Correlation martix}
\label{app2}
\subsection{s-wave scenario}

The Hamiltonian Eq.(\ref{HamS}) for s-wave superconductivity state in graphene 
can be diagonalized by using Bogoluibov transformations
\begin{align}
&e_{\vec{k}+}=\alpha_{+}\frac{1}{\sqrt{2}}(a_{\vec{k},\uparrow}-e^{i \cdot \phi_{\vec{k}}}b_{\vec{k},\uparrow})+
\alpha_{-}\frac{1}{\sqrt{2}}(a_{-\vec{k},\downarrow}^{\dagger}-e^{i \cdot \phi_{\vec{k}}}b_{-\vec{k},\downarrow}^{\dagger}) \label{BogoliubovSCs1}\\
&f_{\vec{k}+}=\beta_{-}\frac{1}{\sqrt{2}}(a_{\vec{k},\uparrow}+e^{i \cdot \phi_{\vec{k}}}b_{\vec{k},\uparrow})-
\beta_{+}\frac{1}{\sqrt{2}}(a_{-\vec{k},\downarrow}^{\dagger}+e^{i \cdot \phi_{\vec{k}}}b_{-\vec{k},\downarrow}^{\dagger}) \label{BogoliubovSCs2}
\end{align}
where
$\alpha_{+}=\sqrt{\frac{1}{2}\left(1+\frac{t \vert \gamma(\vec{k})\vert -\mu}
{\sqrt{\left(t \vert \gamma(\vec{k})\vert - \mu\right)^{2}+\vert 
C_{\vec{k}}\vert^{2}}}\right)}$, 
$\alpha_{-}=\frac{C_{\vec{k}}}{\sqrt{2E_{\alpha}\left(E_{\alpha}+t\vert
\gamma(\vec{k})\vert-\mu\right)}}$,
$\beta_{+}=\sqrt{\frac{1}{2}\left(1+\frac{t \vert \gamma(\vec{k})\vert +\mu}
{\sqrt{\left(t \vert \gamma(\vec{k})\vert + \mu\right)^{2}+\vert 
C_{\vec{k}}\vert^{2}}}\right)}$, and 
$\beta_{-}=\frac{C_{\vec{k}}}{\sqrt{2E_{\beta}\left(E_{\beta}+t\vert
\gamma(\vec{k})\vert+\mu\right)}}$
with
$ E_{\alpha} $ and $ E_{\beta} $ are energies of Bogoliubov quasi-particles
\begin{equation}
E_{\alpha}=\sqrt{\left(t\vert\gamma(\vec{k})\vert-\mu\right)^{2}+\vert C_{\vec{k}}\vert^{2}}
\label{energy_s_alpha}
\end{equation}
and
\begin{equation}
E_{\beta}=\sqrt{\left(t\vert\gamma(\vec{k})\vert+\mu\right)^{2}+\vert C_{\vec{k}}\vert^{2}}.
\label{energy_s_beta}
\end{equation}

The $ e $ ($ f $) sections are determined by Eq.~(\ref{BogoliubovSCs1}) 
(Eq.~(\ref{BogoliubovSCs2})), respectively. These sections are decoupled in 
Bogoliubov description and 
we are allowed than to obtain their contributions to the ground state separative.
We can demand  $ e_{\vec{k}+} \vert G \rangle = 0 $ and 
$ e_{\vec{k}-}^{\dagger} \vert G \rangle = 0 $ where $ \vert G \rangle $
is the ground state. The $ e $ section contributes to the ground state as:
\begin{align}
\prod_{\vec{k}\in IBZ}\left(\alpha_{+}(\vec{k})-
\alpha_{-}(\vec{k})c_{\vec{k}\uparrow}^{\dagger}
c_{-\vec{k}\downarrow}^{\dagger}\right)\vert 0 \rangle
\end{align}
where $ \vert 0 \rangle $ is the vacuum state.
Similar, the contribution of the $ f $ section to the ground state:
\begin{align}
\prod_{\vec{k}\in IBZ}\left(\beta_{-}(\vec{k})+
\beta_{+}(\vec{k})d_{\vec{k}\uparrow}^{\dagger}
d_{-\vec{k}\downarrow}^{\dagger}\right)\vert 0 \rangle
\end{align}
the ground state $ \vert G \rangle $ is determined by conditions:
$ f_{\vec{k}+} \vert G \rangle = 0 $ and 
$ f_{\vec{k}-}^{\dagger} \vert G \rangle = 0 $.
This leads to the complete ground state vector:
\begin{align}
& \prod_{\vec{k}\in IBZ}\left(\alpha_{+}(\vec{k})-
\alpha_{-}(\vec{k})c_{\vec{k}\uparrow}^{\dagger}c_{-\vec{k}\downarrow}^{\dagger}\right)
\nonumber \\
& \prod_{\vec{q}\in IBZ}\left(\beta_{-}(\vec{q})+
\beta_{+}(\vec{q})d_{\vec{q}\uparrow}^{\dagger}d_{-\vec{q}\downarrow}^{\dagger}\right)
\vert 0 \rangle.
\end{align}
Similar findings are obtained for the ground state of the p-wave superconductivity 
state in graphene \cite{Milovanovic11}.

This ground state leads to the correlation matrix when spin $ \downarrow $ is 
traced out:

\begin{equation}
  C(\vec k)=\left(
\begin{array}{cc}
  \frac{1}{2} \left(\vert \alpha_{-}\vert^{2}+\vert \beta_{+}\vert^{2}\right) & \frac{1}{2}e^{-i \phi_{\vec{k}}} \left(\vert \beta_{+}\vert^{2}-\vert \alpha_{-}\vert^{2}\right) \\
  \frac{1}{2}e^{i \phi_{\vec{k}}} \left(\vert \beta_{+}\vert^{2}-\vert \alpha_{-}\vert^{2}\right) & \frac{1}{2} \left(\vert \alpha_{-}\vert^{2}+\vert \beta_{+}\vert^{2}\right)
\end{array}
\label{corrmat}
\right).
\end{equation}

\subsection{chiral d-wave scenario}

Using
\begin{align}
a_{\vec{k}\uparrow}=&-\frac{1}{\sqrt{2}}\left(\alpha_{+}\left(\gamma_{-}^{(1)}
\right)^{\ast}+i\alpha_{-}\gamma_{+}^{(1)}\right)\left(\beta_{+}-\sigma \beta_{-}
\right)o_{\vec{k},+}
-\frac{1}{\sqrt{2}}\left(\alpha_{+}\left(\gamma_{-}^{(2)}\right)^{\ast}+
i\alpha_{-}\gamma_{+}^{(2)}\right)\left(\beta_{+}+\sigma \beta_{-}\right)p_{\vec{k},+} \nonumber \\
&+\frac{1}{\sqrt{2}}\left(\alpha_{+}\gamma_{+}^{(2)}-
i\alpha_{-}\gamma_{-}^{(2)}\right)
\left(\beta_{+}+\sigma \beta_{-}\right)p_{-\vec{k},-}^{\dagger}+
\frac{1}{\sqrt{2}}\left(\alpha_{+}\gamma_{+}^{(1)}-
i\alpha_{-}\gamma_{-}^{(1)}\right)
\left(\beta_{+}-\sigma \beta_{-}\right) o_{-\vec{k},-}^{\dagger}
\end{align} 
we can calculate the mean occupancy at cite A:

\begin{align}
\langle a_{\vec{k}\uparrow}^{\dagger} a_{\vec{k}\uparrow} \rangle=&
\frac{1}{2}
  \left(\alpha_{+}^{2}\vert\gamma_{-}^{(1)}\vert^{2}+
  \vert\alpha_{-}\vert^{2}\left(\gamma_{+}^{(1)}\right)^{2}
  +i\alpha_{+}\gamma_{+}^{(1)}\left(\alpha_{-}\gamma_{-}^{(1)}-\alpha_{-}^{\ast}
  \left(\gamma_{-}^{(1)}\right)^{\ast}\right)\right)
  \left(\beta_{+}-\sigma\beta_{-}\right)^{2} n_{\vec{k}}^{(0)}\nonumber \\
  +&
  \frac{1}{2}
  \left(\alpha_{+}^{2}\vert\gamma_{-}^{(2)}\vert^{2}+
  \vert\alpha_{-}\vert^{2}\left(\gamma_{+}^{(2)}\right)^{2}
  +i\alpha_{+}\gamma_{+}^{(2)}\left(\alpha_{-}\gamma_{-}^{(2)}-\alpha_{-}^{\ast}
  \left(\gamma_{-}^{(2)}\right)^{\ast}\right)\right)
  \left(\beta_{+}+\sigma\beta_{-}\right)^{2} n_{\vec{k}}^{(0)} \nonumber \\
  +&\frac{1}{2}
  \left(\alpha_{+}^{2}(\gamma_{+}^{(1)})^{2}+
  \vert\alpha_{-}\vert^{2}\vert\gamma_{-}^{(1)}\vert^{2}
  -i\alpha_{+}\gamma_{+}^{(1)}\left(\alpha_{-}\gamma_{-}^{(1)}-\alpha_{-}^{\ast}
  \left(\gamma_{-}^{(1)}\right)^{\ast}\right)\right)
  \left(\beta_{+}-\sigma\beta_{-}\right)^{2}(1-n_{\vec{k}}^{(0)})\nonumber \\
  +&
  \frac{1}{2}
  \left(\alpha_{+}^{2}(\gamma_{+}^{(2)})^{2}+
  \vert\alpha_{-}\vert^{2}\vert\gamma_{-}^{(2)}\vert^{2}
  -i\alpha_{+}\gamma_{+}^{(2)}\left(\alpha_{-}\gamma_{-}^{(2)}-\alpha_{-}^{\ast}
  \left(\gamma_{-}^{(2)}\right)^{\ast}\right)\right)
  \left(\beta_{+}+\sigma\beta_{-}\right)^{2}(1-n_{\vec{k}}^{(0)}).
  \label{occ_a}
\end{align}

The average number $ n_{\vec{k}}^{(0)} $ of fermions with momentum $ k $
at temperature $ T=0 $ is $ n_{\vec{k}}^{(0)}=0 $. 

Further, we get the mean occupancy at the cite A

\begin{align}
\langle a_{\vec{k}\uparrow}^{\dagger} a_{\vec{k}\uparrow} \rangle=
  &\frac{1}{2}
  \left(\alpha_{+}^{2}(\gamma_{+}^{(1)})^{2}+
  \vert\alpha_{-}\vert^{2}\vert\gamma_{-}^{(1)}\vert^{2}
  -i\alpha_{+}\gamma_{+}^{(1)}\left(\alpha_{-}\gamma_{-}^{(1)}-\alpha_{-}^{\ast}
  \left(\gamma_{-}^{(1)}\right)^{\ast}\right)\right)
  \left(\beta_{+}-\sigma\beta_{-}\right)^{2}\nonumber \\
  +&
  \frac{1}{2}
  \left(\alpha_{+}^{2}(\gamma_{+}^{(2)})^{2}+
  \vert\alpha_{-}\vert^{2}\vert\gamma_{-}^{(2)}\vert^{2}
  -i\alpha_{+}\gamma_{+}^{(2)}\left(\alpha_{-}\gamma_{-}^{(2)}-\alpha_{-}^{\ast}
  \left(\gamma_{-}^{(2)}\right)^{\ast}\right)\right)
  \left(\beta_{+}+\sigma\beta_{-}\right)^{2}.
  \label{occ_a1}
\end{align}

After basic algebra we find that the correlation matrix obtained by tracing out
spin $ \downarrow $ at $ T=0 $ reads 
\begin{align}
C(\vec k)=&\left(
\begin{array}{cccc}
  C_{11}(\vec{k}) & C_{12}(\vec{k}) \\
 C_{12}^{\ast}(\vec{k}) & C_{22}(\vec{k}) 
\end{array}\right) 
\end{align} 
with
\begin{align}
  C_{11}(\vec k)=&\frac{1}{2}
  \left(\alpha_{+}^{2}(\gamma_{+}^{(1)})^{2}+
  \vert\alpha_{-}\vert^{2}\vert\gamma_{-}^{(1)}\vert^{2}
  -i\alpha_{+}\gamma_{+}^{(1)}\left(\alpha_{-}\gamma_{-}^{(1)}-\alpha_{-}^{\ast}
  \left(\gamma_{-}^{(1)}\right)^{\ast}\right)\right)
  \left(\beta_{+}-\sigma\beta_{-}\right)^{2}\nonumber \\
  +&
  \frac{1}{2}
  \left(\alpha_{+}^{2}(\gamma_{+}^{(2)})^{2}+
  \vert\alpha_{-}\vert^{2}\vert\gamma_{-}^{(2)}\vert^{2}
  -i\alpha_{+}\gamma_{+}^{(2)}\left(\alpha_{-}\gamma_{-}^{(2)}-\alpha_{-}^{\ast}
  \left(\gamma_{-}^{(2)}\right)^{\ast}\right)\right)
  \left(\beta_{+}+\sigma\beta_{-}\right)^{2} \nonumber \\
  =&\frac{1}{2}+\frac{1}{4}\frac{\mu}{\sqrt{\mu^{2}+
  \vert S_{\vec{k}}\vert^{2}}}(\epsilon_{1}+m)\frac{1}{E_{\alpha}}\left(1-\frac{m}{\sqrt{t^{2}\vert \gamma(\vec{k})\vert^{2}+m^{2}}}\right)\nonumber \\
+&  \frac{1}{4}\frac{\mu}{\sqrt{\mu^{2}+
\vert S_{\vec{k}}\vert^{2}}}(\epsilon_{2}+m)\frac{1}{E_{\beta}}\left(1+\frac{m}{\sqrt{t^{2}\vert \gamma(\vec{k})\vert^{2}+m^{2}}}\right),
\end{align}
\begin{align}
  C_{22}(\vec k)=&\frac{1}{2}
  \left(\alpha_{+}^{2}(\gamma_{+}^{(1)})^{2}+
  \vert\alpha_{-}\vert^{2}\vert\gamma_{-}^{(1)}\vert^{2}
  +i\alpha_{+}\gamma_{+}^{(1)}\left(\alpha_{-}\gamma_{-}^{(1)}-\alpha_{-}^{\ast}
  \left(\gamma_{-}^{(1)}\right)^{\ast}\right)\right)
  \left(\beta_{+}+\sigma\beta_{-}\right)^{2}\nonumber \\
  +&
  \frac{1}{2}
  \left(\alpha_{+}^{2}(\gamma_{+}^{(2)})^{2}+
  \vert\alpha_{-}\vert^{2}\vert\gamma_{-}^{(2)}\vert^{2}
  +i\alpha_{+}\gamma_{+}^{(2)}\left(\alpha_{-}\gamma_{-}^{(2)}-\alpha_{-}^{\ast}
  \left(\gamma_{-}^{(2)}\right)^{\ast}\right)\right)
  \left(\beta_{+}-\sigma\beta_{-}\right)^{2} \nonumber \\
 =&\frac{1}{2}+\frac{1}{4}\frac{\mu}{\sqrt{\mu^{2}+
 \vert S_{\vec{k}}\vert^{2}}}(\epsilon_{1}-m)\frac{1}{E_{\alpha}}\left(1+\frac{m}{\sqrt{t^{2}\vert \gamma(\vec{k})\vert^{2}+m^{2}}}\right)\nonumber \\
+&  \frac{1}{4}\frac{\mu}{\sqrt{\mu^{2}+\vert S_{\vec{k}}\vert^{2}}}(\epsilon_{2}-m)\frac{1}{E_{\beta}}\left(1-\frac{m}{\sqrt{t^{2}\vert \gamma(\vec{k})\vert^{2}+m^{2}}}\right),
\end{align}
and
\begin{align}
C_{12}(\vec{k})=&\frac{1}{2}e^{-i\phi_{\vec{k}}}
\left(\alpha_{+}^{2}(\gamma_{+}^{(1)})^{2}
-\vert\alpha_{-}\vert^{2}\vert\gamma_{-}^{(1)}\vert^{2}-
i \alpha_{+}\gamma_{+}^{(1)}\left(\alpha_{-}^{(1)}\gamma_{-}^{(1)}+
\left(\alpha_{-}^{(1)}\right)^{\ast}\left(\gamma_{-}^{(1)}\right)^{\ast}\right)\right)
\left(\beta_{+}^{2}-\beta_{-}^{2}\right)\nonumber \\
-&\frac{1}{2}e^{-i\phi_{\vec{k}}}
\left(\alpha_{+}^{2}(\gamma_{+}^{(2)})^{2}
-\vert\alpha_{-}\vert^{2}\vert\gamma_{+}^{(2)}\vert^{2}-
i \alpha_{+}\gamma_{+}^{(2)}\left(\alpha_{-}^{(1)}\gamma_{-}^{(2)}+
\left(\alpha_{-}^{(1)}\right)^{\ast}\left(\gamma_{-}^{(2)}\right)^{\ast}\right)\right)
\left(\beta_{+}^{2}-\beta_{-}^{2}\right) \nonumber \\
=&\frac{1}{4}e^{-i\phi_{\vec{k}}}\left(\left(\frac{\epsilon_{1}}{E_{\alpha}}-\frac{\epsilon_{2}}{E_{\beta}}\right)-i\frac{{\rm Re}(C_{\vec{k}}){\rm Re}(S_{\vec{k}})+{\rm Im}(C_{\vec{k}}){\rm Im}(S_{\vec{k}})}{\sqrt{\mu^{2}+\
\vert S_{\vec{k}}\vert^{2}}}\left(\frac{1}{E_{\alpha}}-\frac{1}{E_{\beta}}\right)\right)\frac{t\vert \gamma(\vec{k})\vert}{\sqrt{t^{2}\vert \gamma(\vec{k})\vert^{2}+m^{2}}}.
\end{align}
Here, one should notice that $ C_{11}(-\vec{k})=C_{22}(\vec{k}) $ and
$ C_{12}(\vec{k})=\left(C_{12}(-\vec{k})\right)^{\ast} $.

Eigenvectors of the correlation matrix

\begin{align}
q_{\vec{k}\uparrow}=&\delta_{+}(\vec{k})a_{\vec{k}\uparrow}+
\delta_{-}(\vec{k})b_{\vec{k}\uparrow} \\
r_{\vec{k}\uparrow}=&\delta_{+}(-\vec{k})a_{\vec{k}\uparrow}-
\delta_{-}^{\ast}(-\vec{k})b_{\vec{k}\uparrow}
\end{align}

where:

\begin{align}
\delta_{+}(\vec{k})=&\sqrt{\frac{1}{2}\left(1+\frac{C_{11}-C_{22}}{\sqrt{\left(C_{11}-C_{22}\right)^{2}+4\vert C_{12}\vert^{2}}}\right)} \nonumber \\
\delta_{-}(\vec{k})=&\frac{2C_{12}}{\sqrt{2\sqrt{\left(C_{11}-C_{22}\right)^{2}+4\vert C_{12}\vert^{2}}(C_{11}-C_{22}+\sqrt{\left(C_{11}-C_{22}\right)^{2}+4\vert C_{12}\vert^{2}})}} 
\label{delta}
\end{align}

Finally,
we find that the correlation matrix obtained by tracing out
one sublattice, B for example 
\begin{align}
C(\vec k)=&\left(
\begin{array}{cccc}
  C_{11}(\vec{k}) & C_{13}(\vec{k}) \\
 C_{13}^{\ast}(\vec{k}) & C_{33}(\vec{k}) 
\end{array}\right) 
\end{align} 
with
\begin{align}
  C_{11}(\vec k)=&\frac{1}{2}
  \left(\alpha_{+}^{2}(\gamma_{+}^{(1)})^{2}+
  \vert\alpha_{-}\vert^{2}\vert\gamma_{-}^{(1)}\vert^{2}
  -i\alpha_{+}\gamma_{+}^{(1)}\left(\alpha_{-}\gamma_{-}^{(1)}-\alpha_{-}^{\ast}
  \left(\gamma_{-}^{(1)}\right)^{\ast}\right)\right)
  \left(\beta_{+}-\sigma\beta_{-}\right)^{2}\nonumber \\
  +&
  \frac{1}{2}
  \left(\alpha_{+}^{2}(\gamma_{+}^{(2)})^{2}+
  \vert\alpha_{-}\vert^{2}\vert\gamma_{-}^{(2)}\vert^{2}
  -i\alpha_{+}\gamma_{+}^{(2)}\left(\alpha_{-}\gamma_{-}^{(2)}-\alpha_{-}^{\ast}
  \left(\gamma_{-}^{(2)}\right)^{\ast}\right)\right)
  \left(\beta_{+}+\sigma\beta_{-}\right)^{2} \nonumber \\
  =&\frac{1}{2}+\frac{1}{4}\frac{\mu}{\sqrt{\mu^{2}+
  \vert S_{\vec{k}}\vert^{2}}}(\epsilon_{1}+m)\frac{1}{E_{\alpha}}\left(1-\frac{m}{\sqrt{t^{2}\vert \gamma(\vec{k})\vert^{2}+m^{2}}}\right)\nonumber \\
+&  \frac{1}{4}\frac{\mu}{\sqrt{\mu^{2}+
\vert S_{\vec{k}}\vert^{2}}}(\epsilon_{2}+m)\frac{1}{E_{\beta}}\left(1+\frac{m}{\sqrt{t^{2}\vert \gamma(\vec{k})\vert^{2}+m^{2}}}\right),
\end{align}
\begin{align}
  C_{33}(\vec k)=&\frac{1}{2}
  \left(\alpha_{+}^{2}(\gamma_{-}^{(1)})^{2}+
  \vert\alpha_{-}\vert^{2}\vert\gamma_{+}^{(1)}\vert^{2}
  -i\alpha_{+}\gamma_{+}^{(1)}\left(\alpha_{-}\gamma_{-}^{(1)}-\alpha_{-}^{\ast}
  \left(\gamma_{-}^{(1)}\right)^{\ast}\right)\right)
  \left(\beta_{+}+\sigma\beta_{-}\right)^{2}\nonumber \\
  +&
  \frac{1}{2}
  \left(\alpha_{+}^{2}(\gamma_{-}^{(2)})^{2}+
  \vert\alpha_{-}\vert^{2}\vert\gamma_{+}^{(2)}\vert^{2}
  -i\alpha_{+}\gamma_{+}^{(2)}\left(\alpha_{-}\gamma_{-}^{(2)}-\alpha_{-}^{\ast}
  \left(\gamma_{-}^{(2)}\right)^{\ast}\right)\right)
  \left(\beta_{+}-\sigma\beta_{-}\right)^{2} \nonumber \\
 =&\frac{1}{2}-\frac{1}{4}\frac{\mu}{\sqrt{\mu^{2}+
 \vert S_{\vec{k}}\vert^{2}}}(\epsilon_{1}-m)\frac{1}{E_{\alpha}}\left(1+\frac{m}{\sqrt{t^{2}\vert \gamma(\vec{k})\vert^{2}+m^{2}}}\right)\nonumber \\
-&  \frac{1}{4}\frac{\mu}{\sqrt{\mu^{2}+\vert S_{\vec{k}}\vert^{2}}}(\epsilon_{2}-m)\frac{1}{E_{\beta}}\left(1-\frac{m}{\sqrt{t^{2}\vert \gamma(\vec{k})\vert^{2}+m^{2}}}\right),
\end{align}
and
\begin{align}
C_{13}(\vec{k})=&\frac{1}{2}
\left(\alpha_{+}^{2}\left(\gamma_{+}^{(1)}\gamma_{-}^{(1)}-
\gamma_{+}^{(2)}\gamma_{-}^{(2)}\right)
-\left(\alpha_{-}^{\ast}\right)^{2}\left(\gamma_{+}^{(1)}(\gamma_{-}^{(1)})^{\ast}-
\gamma_{+}^{(2)}(\gamma_{-}^{(2)})^{\ast}\right)\right)
\left(\beta_{+}^{2}-\beta_{-}^{2}\right) \nonumber \\
=&\frac{1}{4}\left(\frac{1}{E_{\alpha}}-\frac{1}{E_{\beta}}\right)\frac{\mu}{\sqrt{\mu^{2}+
  \vert S_{\vec{k}}\vert^{2}}}\frac{t\vert \gamma(\vec{k})\vert}{\sqrt{t^{2}\vert \gamma(\vec{k})\vert^{2}+m^{2}}}C_{\vec{k}}^{\ast}.
\end{align}

\end{widetext}
{}


\end{document}